\documentclass[12pt]{article}
 \pdfoutput=1
\usepackage{amssymb}
\usepackage{amsmath}
\usepackage{amstext}
\usepackage{graphicx,epsfig}
\usepackage{epsfig}
\usepackage{verbatim} 
\usepackage{fancyhdr}
\usepackage{fancybox}
\usepackage{color}
\usepackage{ulem,bbold}
\usepackage{enumitem}
\usepackage{subfigure}
\usepackage{bbm}
\usepackage{parskip}

\newcommand{\Comment}[1]{{}}
\definecolor{MyDarkBlue}{rgb}{0.15,0.15,0.45}
\usepackage[linktocpage=true]{hyperref}
\hypersetup{
colorlinks=true,
citecolor=MyDarkBlue,
linkcolor=MyDarkBlue,
urlcolor=MyDarkBlue,
}

\newcommand{\be}{\begin{equation}}  
\newcommand{\ee}{\end{equation}}  
\newcommand{\bea}{\begin{eqnarray}}  
\newcommand{\eea}{\end{eqnarray}}
\newcommand{\nn}{\nonumber}

\newcommand{\s}{{\cal S}} 
\newcommand{\w}{\wedge}

\newcommand{\Eb}{\mathbf E}
\newcommand{\mX}{ X}  
\newcommand{\mY}{ Y}  
\newcommand{\mZ}{ Z}
\newcommand{\mW}{ W}  
\newcommand{\mA}{X}  
\newcommand{\mB}{ Y}  
\newcommand{\mC}{ Z}
\newcommand{\N}{{\cal N}}
\newcommand{\m}{{\rm m}}

\newcommand{\gone}{{g_{^{\!\hspace{.2mm}(1)}}\!}}
\newcommand{\gtwo}{{g_{^{\!\hspace{.2mm}(2)}}\!}}
\newcommand{\gI}{{g_{^{\!\hspace{.2mm}(I)}}\!}}
\newcommand{\gJ}{{g_{^{\!\hspace{.2mm}(J)}}\!}}
\newcommand{\Eone}{{E_{^{\!\hspace{.2mm}(1)}}\!}}
\newcommand{\Etwo}{{E_{^{\!\hspace{.2mm}(2)}}\!}}
\newcommand{\Ethr}{{E_{^{\!\hspace{.2mm}(3)}}\!}}
\newcommand{\Efou}{{E_{^{\!\hspace{.2mm}(4)}}\!}}
\newcommand{\EI}{{E_{^{\!\hspace{.2mm}(I)}}\!}}
\newcommand{\EJ}{{E_{^{\!\hspace{.2mm}(J)}}\!}}
\newcommand{\EIone}{{E_{^{\!\hspace{.2mm}(I_1)}}\!}}
\newcommand{\EItwo}{{E_{^{\!\hspace{.2mm}(I_2)}}\!}}
\newcommand{\EID}{{E_{^{\!\hspace{.2mm}(I_D)}}\!}}
\newcommand{\bEone}{{{\bf E}_{^{\!\hspace{.2mm}(1)}}\!}}
\newcommand{\bEtwo}{{{\bf E}_{^{\!\hspace{.2mm}(2)}}\!}}
\newcommand{\bEthr}{{{\bf E}_{^{\!\hspace{.2mm}(3)}}\!}}
\newcommand{\bEfou}{{{\bf E}_{^{\!\hspace{.2mm}(4)}}\!}}
\newcommand{\bEI}{{{\bf E}_{^{\!\hspace{.2mm}(I)}}\!}}

\newcommand{\bEIone}{{{\bf E}_{^{\!\hspace{.2mm}(I_1)}}\!}}
\newcommand{\bEItwo}{{{\bf E}_{^{\!\hspace{.2mm}(I_2)}}\!}}
\newcommand{\bEID}{{{\bf E}_{^{\!\hspace{.2mm}(I_D)}}\!}}
\newcommand{\barE}{{\bar E}}
\newcommand{\barEone}{{{\bar E}_{^{\!\hspace{.2mm}(1)}}\!}}

\newcommand{\eone}{{e_{^{\!\hspace{.2mm}(1)}}\!}}
\newcommand{\etwo}{{e_{^{\!\hspace{.2mm}(2)}}\!}}
\newcommand{\ethr}{{e_{^{\!\hspace{.2mm}(3)}}\!}}
\newcommand{\efou}{{e_{^{\!\hspace{.2mm}(4)}}\!}}
\newcommand{\eI}{{e_{^{\!\hspace{.2mm}(I)}}\!}}

\newcommand{\de}{ e}

\newcommand{\doteI}{{{\dot e}_{^{\!\hspace{.2mm}(I)}}\!}}

\newcommand{\piI}{{\pi_{^{\!\hspace{.2mm}(I)}}\!}}
\newcommand{\Lone}{{\Lambda_{^{\!\hspace{.2mm}(1)}}\!}}
\newcommand{\Ltwo}{{\Lambda_{^{\!\hspace{.2mm}(2)}}\!}}

\newcommand{\LI}{{\Lambda_{^{\!\hspace{.2mm}(I)}}\!}}

\newcommand{\lI}{{\lambda_{^{\!\hspace{.2mm}(I)}}\!}}

\newcommand{\MI}{{M_{^{\!\hspace{.2mm}(I)}}\!}}
\newcommand{\NI}{{N_{^{\!\hspace{.2mm}(I)}}\!}}
\newcommand{\None}{{N_{^{\!\hspace{.2mm}(1)}}\!}}
\newcommand{\Ntwo}{{N_{^{\!\hspace{.2mm}(2)}}\!}}

\newcommand{\NN}{{N_{^{\!\hspace{.2mm}({\cal N})}}\!}}
\newcommand{\fI}{{f_{^{\!\hspace{.2mm}(I)}}\!}}
\newcommand{\fone}{{f_{^{\!\hspace{.2mm}(1)}}\!}}
\newcommand{\ftwo}{{f_{^{\!\hspace{.2mm}(2)}}\!}}

\newcommand{\pone}{{p_{^{\!\hspace{.2mm}(1)}}\!}}

\newcommand{\pI}{{p_{^{\!\hspace{.2mm}(I)}}\!}}
\newcommand{\gammaI}{{\gamma_{^{\!\hspace{.2mm}(I)}}\!}}

\newcommand{\Cone}{{{\cal C}_{^{\!\hspace{.2mm}(1)}}\!}}

\newcommand{\CI}{{{\cal C}_{^{\!\hspace{.2mm}(I)}}\!}}

\newcommand{\PI}{{{\cal P}_{^{\!\hspace{.2mm}(I)}}\!}}

\def\({\left(}
\def\){\right)}
\def\intx{\int d^Dx\,}

\numberwithin{equation}{section}

\begin{document}


\begin{center}
{\Large \bf{Interacting Spin-2 Fields}}
\end{center} 
 \vspace{1truecm}
\thispagestyle{empty} \centerline{
{\large  { Kurt Hinterbichler${}^{a,}$}}\footnote{E-mail address: \Comment{\href{mailto:kurthi@physics.upenn.edu}}{\tt kurthi@physics.upenn.edu}} 
{\large  {and Rachel A. Rosen${}^{b,}$}}\footnote{E-mail address: \Comment{\href{mailto:rar2172@columbia.edu}}{\tt rar2172@columbia.edu}}
                                                          }

\vspace{1cm}

\centerline{{\it ${}^a$ 
Center for Particle Cosmology, Department of Physics and Astronomy,}}
 \centerline{{\it University of Pennsylvania, Philadelphia, PA 19104, USA}} 
 
 \vspace{1cm}

\centerline{{\it ${}^b$ 
Physics Department and Institute for Strings, Cosmology, and Astroparticle Physics,}}
 \centerline{{\it Columbia University, New York, NY 10027, USA}}

\begin{abstract}
We construct consistent theories of multiple interacting spin-2 fields in arbitrary spacetime dimensions using a vielbein formulation.  We show that these theories have the additional primary constraints needed to eliminate potential ghosts, to all orders in the fields, and to all orders beyond any decoupling limit.  We postulate that the number of spin-2 fields interacting at a single vertex is limited by the number of spacetime dimensions.  We then show that, for the case of two spin-2 fields, the vielbein theory is equivalent to the recently proposed theories of ghost-free massive gravity and bi-metric gravity.  The vielbein formulation greatly simplifies the proof that these theories have an extra primary constraint which eliminates the Boulware-Deser ghost.

\end{abstract}

\newpage

\thispagestyle{empty}
\tableofcontents
\newpage
\setcounter{page}{1}
\setcounter{footnote}{0}

\section{Introduction and summary}
\parskip=5pt
\normalsize

Theories that propagate a massive spin-2 particle have traditionally been plagued by the presence of an additional unstable degree of freedom  -- the Boulware-Deser ghost \cite{Boulware:1973my}.  For the case of a single interacting massive graviton, this problem was solved only recently, by the dRGT (de Rham, Gabadadze, Tolley) massive gravity theories described in \cite{deRham:2010ik,deRham:2010kj}.  These theories were arrived at by choosing interaction terms so as to raise the cutoff of the effective theory \cite{ArkaniHamed:2002sp,Creminelli:2005qk}, after which the Boulware-Deser ghost is automatically vanquished.  (See \cite{Hinterbichler:2011tt} for a recent review of these aspects of massive gravity.)

The dRGT theories can be put in the following form \cite{Hassan:2011vm},
\be 
\s = {M_P^2\over 2}\int d^4x\ \sqrt{-g}\left[R-{m^2\over 4} \sum_{n= 0}^4\beta_n \,S_n(\sqrt{g^{-1}\eta}) \right]\, .
\label{dRGTtheory} 
\ee
This is the Einstein-Hilbert kinetic term for the metric $g_{\mu\nu}$, supplemented by a potential term containing no derivatives of the dynamical metric $g_{\mu\nu}$ but depending explicitly on an external flat metric $\eta_{\mu\nu}$ which breaks the diffeomorphism invariance of the $m=0$ theory\footnote{Of course, the diffeomorphism invariance of massive gravity can be restored by the introduction of St\"{u}ckelberg fields $\pi^A$.  In this work we treat massive gravity in the unitary gauge in which $\pi^A=0$.}.  
Here, $S_n$ is the $n$-th elementary symmetric polynomial (see Appendix \ref{charpolyappendix}) of the eigenvalues of the matrix square root of $g^{\mu\sigma}\eta_{\sigma\nu}$, and the $\beta_n$ are free coefficients.

The dRGT theories \eqref{dRGTtheory} are ghost-free \cite{Hassan:2011hr,deRham:2011rn,deRham:2011qq,Hassan:2011ea,Mirbabayi:2011aa,Golovnev:2011aa,Hassan:2012qv,Kluson:2012wf}.  This means they propagate only the five degrees of freedom of a massive spin-2 field, and not the additional unstable degrees of freedom which plague all interaction terms not of the form of those in \eqref{dRGTtheory} (including those studied in \cite{Boulware:1973my}).  

The counting of degrees of freedom can be done in the Hamiltonian formulation.  As in General Relativity (GR), we adopt ADM variables \cite{Arnowitt:1960es} and Legendre transform with respect to the spatial components of the metric, obtaining a 12 dimensional phase space consisting of the 6 components of the symmetric spatial metric $g_{ij}$, and the 6 components of its canonical momenta $\pi^{ij}$.  In massless GR, the lapse and shift variables appear as Lagrange multipliers, enforcing 4 first class constants which remove 8 degrees of freedom, leaving a 4 dimensional physical phase space which describes the 2 polarizations of the massless graviton and their conjugate momenta.  For generic potentials however, the lapse and shift variables appear as auxiliary variables rather than Lagrange multipliers.  They can therefore be eliminated by their own equations of motion, leaving no constraints, so the entire 12 dimensional phase space is physical, describing the 5 polarizations of the massive graviton and the single Boulware-Deser ghost, as well as their conjugate momenta.  For the special interactions of dRGT massive gravity, however, the lapse still survives as a Lagrange multiplier, enforcing a single primary constraint \cite{Hassan:2011hr}.  This leads to a secondary constraint \cite{Hassan:2011ea} which, together with the primary constraint, forms a second class pair of constraints.  This leaves a 10 dimensional physical space space, just right for the 5 degrees of freedom of  a massive graviton and its canonical momenta.  

The potential of \eqref{dRGTtheory} is awkward, involving matrix square roots.  In particular, this makes it difficult to identify the extra primary constraint which removes the Boulware-Deser ghost.  In the formulation \eqref{dRGTtheory}, this constraint is only seen after a complicated re-definition of the shift variable.  The square root structure suggests that vierbein variables $\Eb^A = E_\mu^{\ A} dx^\mu$ may be better suited for describing this theory since they are, in a sense, like the square root of the metric,
\be 
g_{\mu\nu}=E_\mu^{\ A}E_\nu^{\ B}\eta_{AB} \, .
\ee
In this paper, we will show that the dRGT theory \eqref{dRGTtheory} is precisely equivalent to the following action written in terms of vierbeins,
\begin{multline}
\s =
{M_P^{2}\over 2}\Bigg(\int d^4x\ (\det E) R[E]\\-{m^2\over 4}\int \sum_{n= 0}^4 \frac{\beta_n}{n!(4-n)!} \tilde \epsilon_{A_1A_2A_3 A_4} {\bf 1}^{ A_1}\wedge \cdots\wedge  {\bf 1}^{ A_n}\wedge \Eb^{ A_{n+1}}\wedge \cdots\wedge \Eb^{ A_4}\Bigg) \, , 
\label{drgtviel4}
\end{multline}
where ${\bf 1}^A = \delta_\mu^{\ A} dx^\mu$ is the identity vierbein, which can be thought of as a vierbein for the flat background metric $\eta_{\mu\nu}$.  The flat space epsilon symbol is denoted by $\tilde \epsilon_{A_1A_2\cdots A_4}$.

The potential terms now appear simply as all possible wedge products of the vierbein of the dynamical metric with the vierbein for the background.  The structure of the wedge product is responsible for the appearance of the symmetric polynomials in \eqref{dRGTtheory}.  In these variables, we will find that it is almost trivial to see the existence of the extra primary constraint which makes the theory ghost-free, to all orders in the fields, and to all orders beyond any decoupling limit.

In \eqref{dRGTtheory}, the flat metric $\eta_{\mu\nu}$ is fixed.  One can promote this to a general reference metric, $f_{\mu\nu}$, and the theory is still ghost-free \cite{Hassan:2011tf}.  Going further, one can promote $f_{\mu\nu}$ to a dynamical metric by adding a kinetic term for it,
\be  
\s = \int d^4x \left[{M_{g}^2\over 2} \sqrt{-g}\,R[g]+{M_f^2\over 2} \sqrt{-f}\,R[f]-{m^2 M_{f\!g}^2\over 8}  \sqrt{-g}\, \sum_{n= 0}^4\beta_n \,S_n(\sqrt{g^{-1}f})\right] \, .
\label{dRGTbimetric}
\ee
Unlike the theory with a fixed reference metric, there is now an overall diffeomorphism invariance.  This is a bi-gravity theory, describing at linear level a massless helicity-2 graviton plus a massive spin-2 graviton, for a total of 7 degrees of freedom.  For a general choice of interaction terms, a generic bi-gravity theory would also have a Boulware-Deser-like ghost, for a total of 8 degrees of freedom.  But with the interaction terms of \eqref{dRGTtheory}, it can be shown that the ghost is absent \cite{Hassan:2011zd} so that there are in fact 7 degrees of freedom.

We will show in this paper that the bi-metric theory can also be written simply in vierbein form, 
\begin{multline}
\s =  {M_g^2\over 2} \int d^4x\ \(\det\Eone\) R[\Eone]
+ {M_f^2\over 2} \int d^4x\ \(\det\Etwo\) R[\Etwo]  \\
 -{m^2 M_{f\!g}^2\over 8}  \int  \sum_{n= 0}^4 \frac{\beta_n}{n!(4-n)!} \tilde \epsilon_{A_1A_2A_3 A_4}\bEtwo^{ A_1}\wedge \cdots\wedge \bEtwo^{ A_n}\wedge \bEone^{ A_{n+1}}\wedge \cdots\wedge \bEone^{ A_4}\, ,
\label{bimetricvierbein}
\end{multline}
where $\Eone$ and $\Etwo$ are vierbeins for the two metrics $g_{\mu\nu}$ and $f_{\mu\nu}$, respectively.
In this form, the extra primary constraint responsible for eliminating the Boulware-Deser ghost will be easy to see.

By using the vierbein formulation, we can go beyond the bi-gravity theory and construct ghost-free interaction terms that directly mix more than two gravitons.  These are terms which cannot be easily inferred using the metric formulation.  We can have tri-metric vertices which link together 3 different metrics.  There are three possibilities, depending on which vierbein appears twice in the wedge product,
\bea  
&& \tilde \epsilon_{A_1 A_2 A_3 A_4}\,\bEone^{A_1}\w \bEone^{A_2}\w \bEtwo^{A_3}\w \bEthr^{A_4}\,, \nn\\
 && \tilde \epsilon_{A_1 A_2 A_3 A_4}\,\bEone^{A_1}\w \bEtwo^{A_2}\w \bEtwo^{A_3}\w \bEthr^{A_4}\,, \nn\\
  && \tilde \epsilon_{A_1 A_2 A_3 A_4}\,\bEone^{A_1}\w \bEtwo^{A_2}\w \bEthr^{A_3}\w \bEthr^{A_4}\,.
 \eea
Finally, there is one possible tetra-metric vertex which can mix 4 different vierbeins,
\be
\tilde \epsilon_{A_1A_2A_3A_4} \, \bEone^{A_1}\w \bEtwo^{A_2}\w \bEthr^{A_3}\w \bEfou^{A_4}\,.
\ee
There is no ghost-free vertex of this type which can link 5 or more metrics, since this would require wedging 5 or more vierbeins, which can't be done in 4 dimensions.

Generic theories of $\N$ interacting spin-2 fields will contain the degrees of freedom of $1$ massless spin-2, $\N-1$ massive spin-2's and $\N-1$ Boulware-Deser-like scalar ghosts.  This is because each metric carries 6 potentially propagating degrees of freedom, rather than 5, and there is only one overall diffeomorphism invariance.  In this paper we will show that the above potential terms have the additional primary constraints needed to eliminate the $\N-1$ ghosts, non-linearly to all orders in the fields and to all orders beyond any decoupling limit.

The vierbein formulation and the proof of ghost-freedom can be extended to $D$ spacetime dimensions.  We will thus work in arbitrary dimensions for the remainder of the paper.  The dRGT interaction terms remain the same, but with $D$ symmetric polynomials.  For a theory with multiple interacting gravitons indexed by $I$, our conjecture is that the following term is the most general ghost-free potential in $D$ spacetime dimensions,
\be
U= \sum_{I_1, \ldots,I_D=1}^{\N}\,T^{I_1 I_2\cdots I_D}\, \tilde \epsilon_{A_1A_2\cdots A_D} \, \bEIone^{\ A_1}\w \bEItwo^{\ A_2}\w \ldots\w \bEID^{\ A_D} \, ,
\ee
where $T^{I_1 I_2\cdots I_D}$ is a completely symmetric constant tensor of coefficients.  We will show that this term yields the primary constraints needed to eliminate all of the $\N-1$ ghosts, in all dimensions, and to all orders in the fields and beyond any decoupling limit.  Since these terms are constructed by wedging together combinations of the vielbeins, they are limited by the number of spacetime dimensions.  These ghost-free vertices can directly connect only up to $D$ different gravitons.  

\begin{figure}[h]
\begin{center}
\epsfig{file=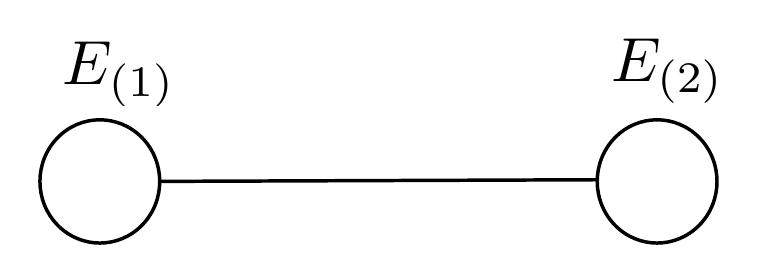,height=.6in,width=1.6in}
\caption{\small \textit {Visual depiction of a bi-metric theory.  The two circles represent the two vielbeins, the bar between them the interaction terms which mix them.}}
\label{bivertex}
\end{center}
\end{figure}

Following the ideas of \cite{ArkaniHamed:2002sp,ArkaniHamed:2003vb}, we can represent these multi-metric theories graphically. The bi-metric theory is given in figure \ref{bivertex}.  The two nodes represent the two metrics, each of which comes with an Einstein-Hilbert term.  The line connecting the nodes represents the bi-metric mass term, which mixes the two metrics.  In the vielbein formalism, each node has a separate diffeomorphism invariance, and a separate local Lorentz invariance (LLI).  The mass term breaks this down to the diagonal subgroup of diffeomorphisms and LLI's, those transformations which act the same way on both metrics together.

\begin{figure}[h]
\begin{center}
\epsfig{file=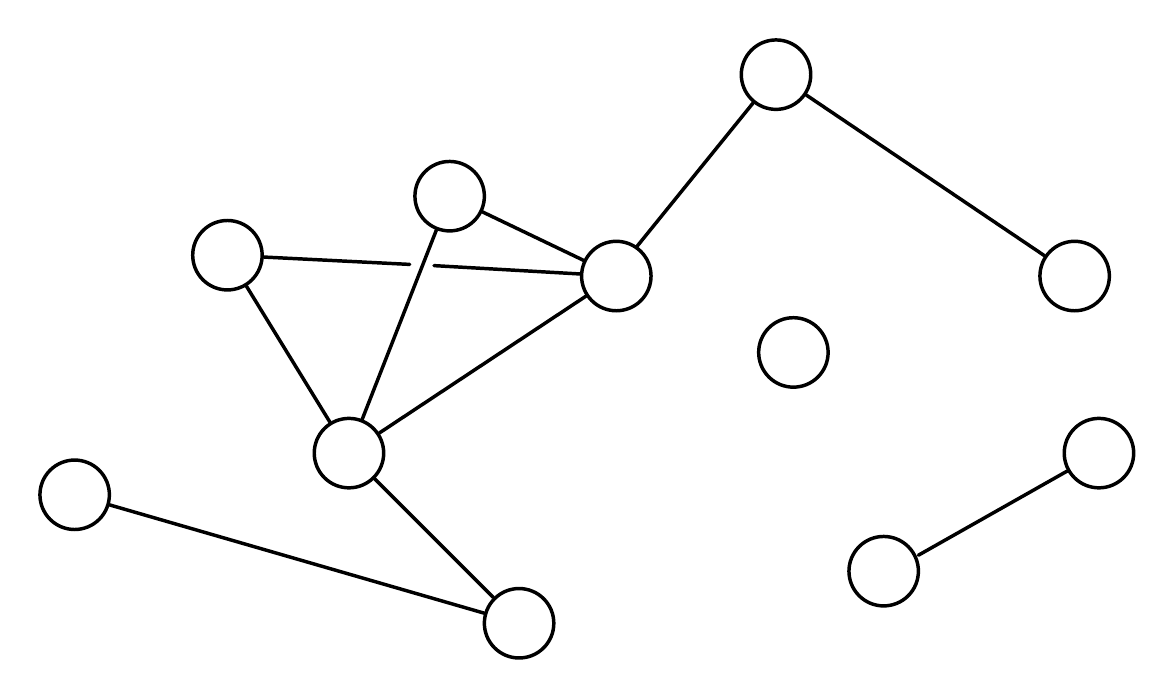,height=2in,width=3.2in}
\caption{\small \textit {Visual depiction of a general multi-metric theory with only bi-metric interactions.}}
\label{generalgraph}
\end{center}
\end{figure}

Continuing in this way, we can link up any number of gravitons using bi-metric interactions, for example as in figure \ref{generalgraph}.  There will be one unbroken diffeomorphism and one LLI for each ``island'' (i.e., disconnected subgraph) in the graph.  The spectrum will consist of one massless graviton for each island, and the rest of the gravitons will be massive. (This does not contradict theorems forbidding multiple interacting massless gravitons \cite{Boulanger:2000rq}, since the islands don't interact with each other.)  We will show that an arbitrary graph, when the links are constructed using the bi-metric interactions of \eqref{bimetricvierbein}, has the primary constraints to make it ghost-free to all orders.  Thus it is now possible to build the kinds of theories in \cite{ArkaniHamed:2002sp}, for deconstructing gravitational dimensions, in a ghost-free manner. 

\begin{figure}[h]
\begin{center}
\epsfig{file=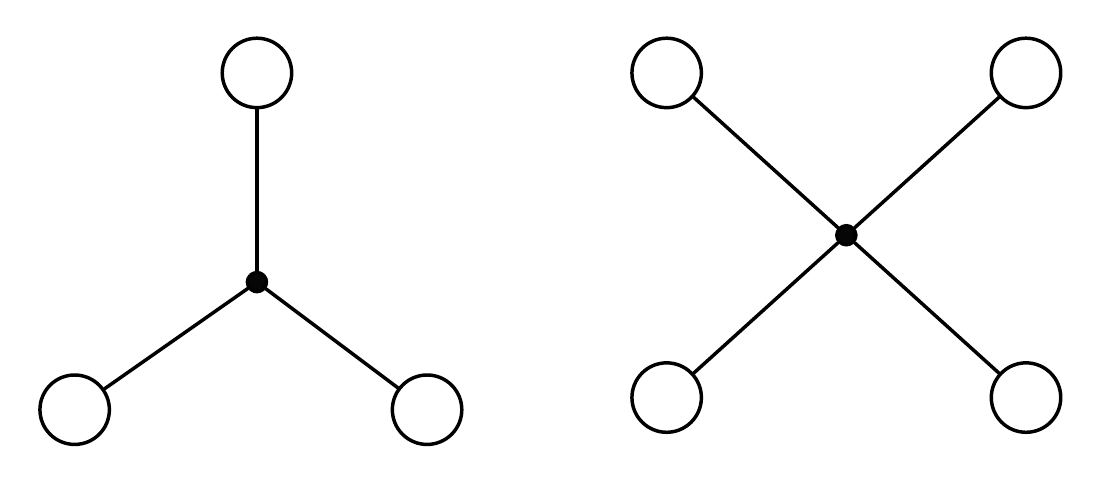,height=1.2in,width=2.5in}
\caption{\small \textit {Visual depiction of the new multi-metric ghost-free interactions discussed in this paper.}}
\label{highervertices}
\end{center}
\end{figure}

The new multi-metric interactions discussed in this paper can be represented as nodes where three or four metrics meet, as in figure \ref{highervertices}.  These can now be used as ingredients in a general theory graph, and the result will still be ghost-free.

Vielbeins have proven useful in the study of massive gravity.  The vielbein formalism within the context of massive gravity has been developed in \cite{Nibbelink:2006sz,Hanada:2008hs,Chamseddine:2011mu,Chamseddine:2011bu,Mirbabayi:2011aa}.  Here we suggest that they may very well be essential.  As we'll see, there is no straightforward way to write down an equivalent metric formulation for theories including these tri-metric or tetra-metric vertices, or even theories with only bi-metric interactions whose theory graphs contain closed loops.  The vielbein theories provide a new and very wide parameter space in which to search for theories that may address issues such as superluminality \cite{Adams:2006sv,Deser:2012qx} and the low cutoff of the effective field theory \cite{ArkaniHamed:2002sp}.

In this paper our analysis will proceed from the general to the specific.  We first introduce the general interaction terms in vielbein form and perform the Hamiltonian analysis.  We show that the vielbein formulation makes it straightforward to identity the primary constraints which eliminate the potential ghosts.  We then consider the recently proposed ghost-free bi-gravity and massive gravity theories which were formulated in terms of metrics rather than vielbeins.  We show that these theories are dynamically equivalent to the vielbein multi-gravity theories, for the case of two metrics.   We repeat the Hamiltonian analysis for the case of massive gravity in the vielbein formulation, addressing the subtleties for this case.

\bigskip
{\bf Conventions}:   
$D$ refers to the number of spacetime dimensions while $d\equiv D-1$ is the number of spatial dimensions.  We use Greek letters $\mu$, $\nu$, etc., for spacetime indices and $i$, $j$, etc., to denote their spatial components.  These are raised and lowered with the full metric, $g_{\mu\nu}$ and the spatial metric $g_{ij}$, respectively.  We use capital  letters $A$, $B$, etc., for Lorentz indices and lower case letters $a$, $b$, etc., for their spatial components.  These are raised and lowered with the full Minkowski metric, $\eta_{AB}$ and the flat spatial metric $\delta_{ab}$, respectively.  The indices $(I)$, $(J)$, etc., label the $\N$ spin-2 fields.  The Einstein summation convention is used for all indices but the $(I)$, $(J)$, etc.  We use the mostly plus metric signature convention, $\eta_{\mu \nu} = (-,+,+,+,\ldots)$.  $E$ denotes the $D$-dimensional vielbein while $e$ denotes the spatial vielbein. Indices are (anti-)symmetrized with weight 1, i.e. $[\mu\nu]={1\over 2}(\mu\nu-\nu\mu).$  The epsilon \textit{symbol} has a tilde, and is always defined so that $\tilde\epsilon_{012\cdots D}=+1$, regardless of the placement of indices or the signature of the metric (so $\tilde\epsilon^{012\cdots D}=+1$ as well).  

\section{Multi-metric theory}

We start with the general case of multiple interacting spin-2 fields in $D$ dimensions.  We have a collection of $\N$ metrics $\gI_{\mu\nu}$, indexed by $I$, each with its own Einstein-Hilbert term and Planck mass $\MI$,
\be
 {\cal L}=\sum_I {\MI^{D-2}\over 2}\sqrt{-\gI}\,R[\gI] \, .
 \ee
Without interaction terms, this action has $\N$ separate diffeomorphism invariances, one for each $I$, given by arbitrary functions $\fI^\mu(x)$,
\be
\label{GRfullgauge}
 \gI_{\mu\nu}(x)\rightarrow {\partial \fI^\alpha \over \partial  x^\mu}{\partial \fI^\beta \over \partial  x^\nu}\gI_{\alpha\beta}\(\fI(x)\) \,.
 \ee

As we wish to express the interactions in vielbein form,  for each metric $\gI_{\mu\nu}(x)$ we introduce a vielbein one-form $\bEI^A=\EI_\mu^{\ A}(x)dx^\mu$, which relates the metric to the flat Minkowski metric $\eta_{AB}$,
\be
\gI_{\mu\nu} = \EI_\mu^{\ A} \, \EI_\nu^{\ B}\,\eta_{AB} \,  . 
\label{gEE}
\ee
We denote the inverse vielbein by $\EI^\mu_{\ A}(x)$, so that $\EI^\mu_{\ A}\EI_\mu^{\ B}=\delta_A^B$ and $\EI^\mu_{\ A}\EI_\nu^{\ A}=\delta_\nu^\mu$.

Without interaction terms, we now have, in addition to the diffeomorphism invariances, a local Lorentz symmetry for each $I$,
\be 
\EI_\mu^{\ A}\rightarrow \LI^{A}_{\ B} \EI^{\ B}_\mu\, ,
\ee
where $\LI^{A}_{\ B}(x)\in SO(1,D-1)$.  The inverse transforms as $\EI^\mu_{\ A}\rightarrow \LI^{\ B}_{A} \EI^\mu_{\ B}$, where we define the inverse $\LI_A^{\ B}\equiv\left(\LI^{-1}\right)^{B}_{\ A}$ so that $\LI_C^{\ A}\LI^C_{\ B}=\LI^A_{\ C}\LI_B^{\ C}=\delta^A_B.$
The vielbeins transform as one-forms under the local diffeomorphism symmetries,
\be 
\EI_{\mu}^{\ A}(x)\rightarrow {\partial \fI^\nu \over \partial  x^\mu}\EI_{\nu}^{\ A}\left(\fI(x)\right) \, .
\ee

Note that, in pure GR, the passage from metric to vielbein via the replacement \eqref{gEE} is nothing but the St\"ukelberg trick:  the metric is symmetric and so has $D(D+1)/2$ components, while the vielbein has no restriction other than invertibility and so has $D^2$ components.  We have thus introduced $D(D-1)/2$ new, unphysical gauge invariances (the local $SO(1,D-1)$ symmetries) along with $D(D-1)/2$ new unphysical fields.

\subsection{The interaction terms}
Our goal is to write down the most general interactions among these spin-2 fields which do not introduce ghosts.  We look for interactions which would be most relevant at long distances, those involving no derivatives of the metric or vielbein.  In addition, we demand that at least one overall diffeomorphism and one overall local Lorentz symmetry remain unbroken, that is, the one for which the $\fI^\mu(x)$ are equal for all $I$, and the $\LI^A_{\ A'}(x)$ are equal for all $I$.  This requires the potential to be a diffeomorphism and Lorentz scalar constructed out of the vielbeins.

Our conjecture is that the following term is the most general ghost-free potential of $\N$ spin-2 fields,
\be
U = \sum_{I_1, \ldots,I_D=1}^{\N} \, T^{I_1 I_2\cdots I_D}\, \tilde \epsilon_{A_1A_2\cdots A_D} \, \bEIone^{\ A_1}\w \bEItwo^{\ A_2}\w \ldots\w \bEID^{\ A_D} \, ,
\label{V}
\ee
where $T^{I_1 I_2\cdots I_D}$ is a completely symmetric constant tensor of coefficients, and $\tilde \epsilon_{A_1A_2\cdots A_D}$ is the totally anti-symmetric flat space epsilon symbol.  This term is invariant under an overall local Lorentz transformation and transforms as a scalar density of weight $1$ under general coordinate transformations.  In $D$ dimensions, the total number of different terms in this $\N$-metric interaction is $\left(\begin{array}{c}\N+D-1 \\D\end{array}\right)$, the number of independent components of a rank $D$ symmetric tensor in $\N$ dimensions.  

Let's look at some of the possible terms.  The simplest case is when there is only one metric $g_{\mu\nu}$, and hence one vielbein $E_\mu^{\ A}$.  Then there is only one possible term,
\be 
\tilde \epsilon_{A_1A_2\cdots A_D}\,{\bf E}^{A_1}\w {\bf E}^{ A_2}\w \ldots\w {\bf E}^{A_D} \, ,
\ee
corresponding to a cosmological constant, $\propto \det E\propto\sqrt{-g}$.

If there are two metrics $\gone$ and $\gtwo$, in addition to a cosmological term for each metric separately, there are now $D-1$ possible terms which mix the two,
\bea  &&  \tilde \epsilon_{A_1 A_2 A_3\cdots A_D}\, \bEtwo^{A_1}\w \bEtwo^{A_2}\w \bEtwo^{A_3}\w \ldots\w \bEtwo^{A_D} \,, \nn\\
&& \tilde  \epsilon_{A_1 A_2 A_3\cdots A_D}\, \bEone^{A_1}\w \bEtwo^{A_2}\w \bEtwo^{A_3}\w \ldots\w \bEtwo^{A_D}\,, \nn\\
&& \tilde  \epsilon_{A_1 A_2 A_3\cdots A_D}\, \bEone^{A_1}\w \bEone^{A_2}\w \bEtwo^{A_3}\w \ldots\w \bEtwo^{A_D}\,, \nn\\
&&\vdots \nn\\
&& \tilde  \epsilon_{A_1 A_2 A_3\cdots A_D}\, \bEone^{A_1}\w \bEone^{a_2}\w \bEone^{A_3}\w \ldots\w \bEone^{A_D}\,.
\label{twometricterms}
\eea
We will see that these are equivalent to the interaction terms of ghost-free bi-gravity, and when one of the metrics is non-dynamical, ghost-free massive gravity.

With multiple metrics, we are allowed to wedge together all the various metrics in all possible ways.  Let us enumerate all the possibilities for $D=4$.  With one metric, there is only the cosmological constant term.  For two metrics, there are the $5$ terms listed in (\ref{twometricterms}); the two cosmological terms for each metric separately plus the three mixing terms.  For three metrics, there are 3 cosmological terms, 3 terms mixing each of the 3 possible pairs of metrics (for a total of 9 bi-metric terms) and there are 3 terms that mix together all three of the metrics, 
\bea  && \tilde  \epsilon_{A_1 A_2 A_3 A_4}\,\bEone^{A_1}\w \bEone^{A_2}\w \bEtwo^{A_3}\w \bEthr^{A_4}\,, \nn\\
 && \tilde  \epsilon_{A_1 A_2 A_3 A_4}\,\bEone^{A_1}\w \bEtwo^{A_2}\w \bEtwo^{A_3}\w \bEthr^{A_4}\,, \nn\\
  &&\tilde  \epsilon_{A_1 A_2 A_3 A_4}\,\bEone^{A_1}\w \bEtwo^{A_2}\w \bEthr^{A_3}\w \bEthr^{A_4}\,,
 \eea
for a grand total of 15 terms.
For 4 metrics, there are 4 cosmological terms, 3 terms mixing each of the 6 possible pairs of metrics (for a total of 18 bi-metric terms), 3 terms that mix together all three metrics in each of the four 3-metric subsets (for a total of 12 tri-metric terms), and finally one term which mixes all four metrics,
\be
\tilde \epsilon_{A_1A_2A_3A_4} \, \bEone^{A_1}\w \bEtwo^{A_2}\w \bEthr^{A_3}\w \bEfou^{A_4}\,,
\ee
for a grand total of 35 terms.  With 5 or more metrics, there are no terms which mix together all of the metrics, so there are only the terms which mix 4 or fewer, for each subset of 4.  In general, the total number of terms in an $\N$-metric theory in 4 dimensions, is $\left(\begin{array}{c}\N+4-1 \\4\end{array}\right)$, which is the number of independent components of a rank 4 symmetric tensor in $\N$ dimensions.  

For general dimension $D$, with $\N$ metrics, the number of possible terms which mix together $n$ of the metrics ($n\leq \N$) can be written 
\be 
\left(\begin{array}{c}\N \\ n\end{array}\right) \left(\begin{array}{c}D-1 \\ n-1\end{array}\right)\,.
\ee
The first factor is the number of ways of choosing the subset of $n$ metrics which are coupled together, and the second factor is the number of ways of partitioning the $D$ terms in the wedge product among the $n$ metrics.  Summing over $n=1\cdots \N$, we find a total of $\left(\begin{array}{c}\N+D-1 \\D\end{array}\right)$ possible terms.  Note that if $\N>D$, there are no terms which mix together all the metrics.

We can graphically represent a theory with $\N$-metrics and the interaction term \eqref{V} by drawing a ``theory graph," (or gravitational quiver diagram) following \cite{ArkaniHamed:2002sp,ArkaniHamed:2003vb,Schwartz:2003vj,Hanada:2010js}.  For each metric, we draw a dot.  For each term in the interaction potential \eqref{V} involving only two of the metrics, we draw a line connecting the corresponding two dots.  For the terms involving three or more metrics, we draw a cubic, quartic, etc. vertex as in Figure \ref{highervertices}, and connect its edges to the corresponding metrics.   The number of unbroken gauge and local Lorentz invariances is the number of disconnected ``islands" in the theory graph.

It will sometimes be more convenient to work with a matrix expression for the the interaction terms, rather than the wedge products.  Accordingly, we derive the following useful result
\bea
&& \tilde \epsilon_{A_1A_2 \cdots A_D}\bEIone^{A_1} \w  \bEItwo^{A_2} \w \ldots \w \bEID^{A_D}  \nonumber \\
&&= \(\det \EIone\, d^Dx \)\,
\tilde \epsilon_{A_1 A_2 \cdots A_D} \,\tilde \epsilon^{B_1 B_2 \cdots B_D}\,
 \delta_{B_1}^{\ A_1} \, (\EIone^{-1} \EItwo)_{B_2}^{\ A_2} \cdots (\EIone^{-1} \EID)_{B_D}^{\ A_D} \,  \nonumber\\
 && =\(\det \EIone\, d^Dx \) S({\bf 1},\EIone^{-1} \EItwo,\ldots, \EIone^{-1} \EID),
 \label{dens}
\eea
 where $(\EIone^{-1} \EItwo)_{B_2}^{\ A_2} = \EIone^{\mu}_{\ B_2}\, \EItwo_{\mu}^{\ A_2}$ and $d^Dx=dx^1 \w dx^2 \w \ldots \w dx^D$ is the volume element, and $S$ is the multi-matrix symmetric polynomial defined and discussed in Appendix \ref{charpolyappendix3}.  

To see this, start by writing the wedge products in terms of epsilon symbols and the volume element,
\be  
 \tilde \epsilon_{A_1A_2 \cdots A_D}\bEIone^{A_1} \w  \bEItwo^{A_2} \w \ldots \w \bEID^{A_D}
 =  \tilde \epsilon_{A_1 A_2 \cdots A_D} \, \tilde  \epsilon^{\mu_1 \mu_2 \cdots \mu_D} \, 
 \EIone_{\mu_1}^{\ A_1} \,  \EItwo_{\mu_2}^{\ A_2} \cdots \EID_{\mu_D}^{\ A_D}  \, d^Dx \,.
 \label{density}
\ee
Then single out $\EIone$ and re-express the curved space epsilon symbol in terms of the flat space epsilon symbol, the determinant $\det \EIone$ and a bunch of inverse vielbeins $\EIone_{\ A}^{\mu}$,
\bea
&&\( \det \EIone\, d^Dx \) \,
 \tilde  \epsilon_{A_1 A_2 \cdots A_D} \, \tilde  \epsilon^{B_1 B_2 \cdots B_D}\,
 \EIone_{\mu_1}^{\ A_1} \,  \EItwo_{\mu_2}^{\ A_2} \cdots \EID_{\mu_D}^{\ A_D}
 \,\EIone_{\ B_1}^{\mu_1} \,  \EIone_{\ B_2}^{\mu_2} \cdots \EIone_{\ B_D}^{\mu_D}   \nonumber\\
&& =\( \det \EIone\, d^Dx \)\,
 \tilde  \epsilon_{A_1 A_2 \cdots A_D} \,  \tilde \epsilon^{B_1 B_2 \cdots B_D}\,
 \delta_{B_1}^{\ A_1} \, (\EIone^{-1} \EItwo)_{B_2}^{\ A_2} \cdots (\EIone^{-1} \EID)_{B_D}^{\ A_D} \, .
\eea
This expression singles out one of the vielbeins to be in the determinant, however, this choice is arbitrary -- there are other equivalent expressions depending on which vielbein is chosen to be in the determinant.
 
As an example, consider the interaction terms in $D=4$.  Define the matrices
\be
\mA  =\Eone^{-1} \Etwo \, , ~~~~\mB  =\Eone^{-1} \Ethr \, , ~~~~\mC  =\Eone^{-1} \Efou \, .
\ee
The ghost-free potentials take the form
\bea
\!\!\!\!\!\!\!\!  \tilde \epsilon_{A_1 A_2 A_3 A_4}\,\bEone^{A_1}\w \bEone^{A_2}\w \bEone^{A_3}\w \bEtwo^{A_4}
 &\!\!\!\!=\!\!\!\!&6\,\( \det \EIone\, d^Dx \) \,[\mA] \, ,\nonumber\\
\!\!\!\!\!\!\!\!  \tilde \epsilon_{A_1 A_2 A_3 A_4}\,\bEone^{A_1}\w \bEone^{A_2}\w \bEtwo^{A_3}\w \bEtwo^{A_4}
 &\!\!\!\!=\!\!\!\!&2\,\( \det \EIone\, d^Dx \) \, \Big([\mA]^2-[\mA^2]\Big) \, ,\nonumber\\
\!\!\!\!\!\!\!\!   \tilde \epsilon_{A_1 A_2 A_3 A_4}\,\bEone^{A_1}\w \bEone^{A_2}\w \bEtwo^{A_3}\w \bEthr^{A_4}
 &\!\!\!\!=\!\!\!\!&2\,\( \det \EIone\, d^Dx \) \, \Big([\mA][\mB]-[\mA\mB]\Big)\, , \nonumber\\
\!\!\!\!\!\!\!\!  \tilde \epsilon_{A_1 A_2 A_3 A_4}\,\bEone^{A_1}\w \bEtwo^{A_2}\w \bEtwo^{A_3}\w \bEtwo^{A_4}
 &\!\!\!\!=\!\!\!\!&\( \det \EIone\, d^Dx \) \, \Big([\mA]^3-3[\mA][\mA^2] +2[\mA^3]\Big) \, ,\nonumber\\
\!\!\!\!\!\!\!\!  \tilde \epsilon_{A_1 A_2 A_3 A_4}\,\bEone^{A_1}\w \bEtwo^{A_2}\w \bEthr^{A_3}\w \bEfou^{A_4}
 &\!\!\!\!=\!\!\!\!&\( \det \EIone\, d^Dx \) \,\Big([\mA][\mB][\mC]- [\mA][\mB\mC]\nonumber\\
 && -[\mB][\mA\mC] -[\mC][\mA\mB] +[\mA\mB\mC]+[\mA\mC\mB]\Big) \, ,\nonumber \\
\eea
as well as every non-redundant permutation of $\eone$, $\etwo$, $\ethr$ and $\efou$.  Along with the four cosmological constants, these give the 35 interaction terms described in the previous section.

\subsection{Hamiltonian formulation \label{hamform}}
In this section we perform a Hamiltonian analysis of the multi-vielbein theory with a general interaction term \eqref{V}.  To this end, we perform a $d+1$ decomposition of the vielbein into canonically conjugate ADM variables.  

A general vielbein can always, by a local Lorentz transformation, be put into upper triangular form (upper triangular vielbeins will be written with a hat), 
\be 
\label{uppertrianvielbein}
\hat{E}_\mu^{\ A}=\left(\begin{array}{cc}N & N^ie_i^{\ a} \\0 & e_i^{\ a}\end{array}\right),\ \ \  \hat{E}^\mu_{\ A}=\left(\begin{array}{cc}{1\over N} & 0 \\-{N^i\over N} & e^i_{\ a}\end{array}\right)\,.
\ee
Here the $N$ and $N^i$ are the $D$ time-like components.  The spatial vielbeins $e_i^{\ a}$ contain $(D-1)^2$ components and are related to the spatial part of the metric by $g_{ij} =e_i^{\ a}e_j^{\ b}\delta_{ab}$. By writing out the metric of this vielbein, we see that $N$ and $N^i$ are the usual lapse and shift of the metric ADM decomposition \cite{Arnowitt:1960es},
\be 
g_{\mu\nu}=\hat E_\mu^{\ A} \hat E_\nu ^{\ B}\eta_{AB}=\left(\begin{array}{cc}-N^2+N^i N_i & N_i\\ N_j & g_{ij} \end{array}\right)\,.
\label{gADM}
\ee
   
The upper triangular form does not completely fix the local Lorentz invariance.  It leaves a residual local spatial rotation.  There are $D$ components in the $N$, $N^i$ and $(D-1)^2$ in the spatial vielbein.   The remaining $D-1$ components of the general vielbein have been fixed by using the upper triangular gauge choice.

We can formulate an arbitrary vielbein as the action of some standard boost on an upper triangular vielbein\footnote{This is analogous to the standard boost used to define single particle states in Lorentz invariant quantum theory.  See for instance chapter 2 of \cite{Weinberg:1995mt}.}.  For every given d-vector $p^a$, we define a standard Lorentz boost
\be
 \Lambda(p)^A_{\ B}=\left(\begin{array}{cc}\gamma & p^a \\p_b & \delta_{\ b}^{a}+\frac{1}{\gamma+1}p^a p_b\end{array}\right) \,,
 \ee
where indices on $p_a$ are raised and lowered with $\delta_{ab}$ and 
\be \gamma \equiv \sqrt{1+p_a p^a}.\ee  
This standard boost takes the standard time-like $D$-vector $(1,0,0,\ldots)$ into the unit normalized $D$-vector with spatial components given by $p^a$,
\be
\Lambda(p)^A_{\ B} \left(\begin{array}{c}1 \\\vec 0\end{array}\right)^B=\left(\begin{array}{c}\gamma \\p^a\end{array}\right)^A \,.
\ee
A general vielbein can now be written as the standard boost of an upper triangular vielbein
 \be
 \label{uptrianparam}
E_\mu^{\ A} =\Lambda(p)^A_{\ B} \hat E_\mu^{\ B} =
\left(\begin{array}{cc}N \gamma +N^i e_i^{\ a} p_a& N p^a+N^i e_i^{\ b} (\delta_b^{\ a}+\frac{1}{\gamma+1}p_b p^a) \\
e_i^{\ a} p_a &  e_i^{\ b} (\delta_b^{\ a}+\frac{1}{\gamma+1}p_b p^a)\end{array}\right) \,.
 \ee 
This is simply a reparametrization of a general vielbein, one which will be particularly convenient for the Hamiltonian analysis.  There need not be any gauge or Lorentz invariance to do this.   The $D^2$ components of the general vielbein are now parametrized by the $D$ components of $N$ and $N^i$,  the $(D-1)^2$ components of the spatial vielbein $e_i^{\ a}$, and the $D-1$ components $p^a$.  
 
We now express the Einstein-Hilbert term in terms of this decomposition.  The Einstein-Hilbert term is invariant under local Lorentz transformations.  Therefore, when we plug in the vielbein as parametrized in \eqref{uptrianparam}, all the $p^a$ dependence drops out.  Thus we can evaluate the Einstein-Hilbert action using the upper triangular ansatz \eqref{uppertrianvielbein}.  

The Hamiltonian formulation of GR in upper triangular vielbein form is reviewed in Appendix \ref{GRveilbeinham}.  The result is that the Einstein-Hilbert kinetic term can be written in the form of a constrained Hamiltonian system on the $2d^2$ dimensional phase space consisting of the spatial vielbein components $e_i^{\ a}$, and their canonical momenta $\pi^i_{\ a}$,
\bea
\s_{EH}=\intx  \(\pi^i_{\ a}\dot e_i^{\ a}-N{\cal C}-N^i{\cal C}_i-{1\over 2}\lambda^{ab}\,{\cal P}_{ab}\)\, .
\eea
Here ${\cal C}(\de,\pi),\  {\cal C}_i(\de,\pi)$ are the usual diffeomorphism constraints of GR, whose Lagrange multipliers are the lapse and shift.   They depend only on the spatial vielbeins and their conjugate momenta.  In addition, we have $d(d-1)/2$ primary constraints $ {\cal P}_{ab}(\de,\pi)=e_{i[a}\pi^i_{\ b]} $, responsible for generating the residual spatial local Lorentz rotations of the upper triangular vielbein, along with their Lagrange multipliers $\lambda^{ab}$.

For a theory with $\N$ spin-2 fields, each spatial vielbein $\eI_i^{\ a}$ gets canonical momenta $\piI^i_{\ a}$.  Each of the Einstein-Hilbert terms is separately Lorentz invariant, and so will not depend on the $\pI^a$'s,
\bea\label{NhamiltonianEH}
\s_{EH}=\intx \  \sum_{I=1}^\N \( \piI^i_{\ a}\doteI_i^{\ a}-\NI \, \CI-\NI^i\, \CI_i-{1\over 2}\lI^{ab}\,\PI_{ab}\) \, .
\eea
Here $\CI(e,\pi)$, $\CI_i(e,\pi)$, are the diffeomorphism constraints of GR, one for each of the diffeomorphisms of the $\N$ Einstein-Hilbert terms, whose Lagrange multipliers are the lapses and shifts.  In addition, there are the $\N$ sets of additional primary constraints, 
\bea \nn 
\PI_{ab}(e,\pi)=\eI_{i[a}\piI^i_{\ b]}, 
\eea
responsible for generating spatial local Lorentz rotations of each of the spatial vielbeins, along with their Lagrange multipliers $\lI^{ab}$.  In the Hamiltonian \eqref{NhamiltonianEH}, the spatial vielbein is unconstrained, and there is a spatial local Lorentz rotation left over as a gauge symmetry, enforced by the primary constraints $\PI_{ab}$.
 
Let's now consider adding a general interaction term \eqref{V}.  We assume the theory graph for our interaction term is connected.  This is no loss of generality, since the various connected islands of a disconnected graph do not interact with each other and can be treated independently.  The proposed potential terms thus have only one overall Lorentz invariance so we are not free to choose a gauge for each vielbein.   For each vielbein we use the general parametrization \eqref{uptrianparam},
\be
\EI_\mu^{\ A}  =
\left(\begin{array}{cc}\NI \gammaI +\NI^i \eI_i^{\ a} \pI_a& \NI \,\pI^a+\NI^i \eI_i^{\ b} (\delta_b^{\ a}+\frac{1}{\gammaI+1}\pI_b \pI^a) \\
\eI_i^{\ a} \pI_a & \eI_i^{\ b} (\delta_b^{\ a}+\frac{1}{\gammaI+1}\pI_b \pI^a) \end{array}\right) \,.
\label{EI}
 \ee 
We may use the overall local Lorentz invariance to set one of the $p^a$'s to zero, say $\pone^a=0$.  This leaves a residual overall local spatial rotation invariance\footnote{In previous versions of this paper, we had proposed an alternative method of dealing with the local spatial rotation invariance using constrained spatial vielbeins.  It's not clear this method works because solving the constraint (eq. 2.27 of v2) may introduce dependence on the lapse or shift into the vielbeins (we thank Shuang-Yong Zhou for pointing this out).}.

\subsection{Ghost-freedom}
We now demonstrate the existence of $\N-1$ Hamiltonian constraints of the $\N$-vielbein theory.  This will guarantee the right number of degrees of freedom to describe one massless graviton and $\N-1$ massive gravitons in $D$ dimensions.

To identify the primary constraints, let us consider now the interaction term \eqref{V}, with the vielbeins given by the parametrization \eqref{EI}.  We observe the crucial fact that both $\EI_0^{\ 0}$ and $\EI_0^{\ b}$ are linear in the lapses $\NI$ and shifts $\NI^i$, while $\EI_i^{\ 0}$ and $\EI_i^{\ b}$ are independent of the lapses and shifts.  Thus, due to the structure of the epsilon tensor, the interaction terms (\ref{V}) are manifestly linear in all the lapses and shifts and can be written in the form
\be
U = \sum _{I=1}^\N  \(  \NI\, \CI^\m + \NI^i\,\CI^\m_i \)\, .
\ee
where $\CI^\m(e,p), \ \CI^\m_i(e,p)$ are functions of the spatial vielbeins $\eI_i^{\ a}$ and the boost vectors $\pI^a$.
The ``m" superscript denotes that these terms are coming from the mass (interaction) term.  

The full action is given by
\be
\s = \intx \sum _{I=1}^\N \left(\piI^i_{\ a} \doteI_i^{\ a}-\NI\left[\CI+\CI^\m\right]-\NI^i\left[\CI_i+\CI^\m_i\right]-{1\over 2}\lI^{ab}\PI_{ab} \right) \, .
\ee
We now use $\N-1$ of the $\N$ shift equations of motion to solve for the $\N-1$ remaining $\pI_a$ variables (recall that we have used the overall local Lorentz invariance to rotate away ${p_{(1)}}^a$),
 \be 
\CI_i(e,\pi)+\CI^\m_i(e,p)=0\ \ \Rightarrow \ \ \pI^a=\pI^a(e,\pi)\, ,
\ \ \ \ \ I=2,\ldots,\N\,.
 \ee
We have thus eliminated all the $\pI$ from the action in favor of the $\eI$ and $\piI$.  After doing this, the action remains linear in the ${\cal N}$ lapse variables $\NI$ and the one remaining shift variable, $\None^i$,
\be
\s = \intx \left\{\sum _{I=1}^\N \left(\piI^i_{\ a} \doteI_i^{\ a}-\NI\left[\CI+\CI^\m\right]-{1\over 2}\lI^{ab}\PI_{ab} \right) -\None^i\left[\Cone_i+\Cone^\m_i\right] \right\}\, .
\ee

There is a lapse and shift constraint with $\None$ and $\None^i$ as multipliers, which will (after mixing with other constraints) generate the overall unbroken diffeomorphism symmetry of the theory.  There are $\N \times \frac{1}{2} d (d-1)$ spatial local  Lorentz (i.e, rotation) constraints.  One combination, the overall sum $\sum_I {{\cal P}_{(I)}}_{ab}$, will generate the unbroken overall spatial local Lorentz invariance.  The other combinations will lead to secondary constraints.  

Finally, and most importantly, we see the presence of the additional $\N-1$ primary constraints with $\Ntwo,\ldots, \NN$ as multipliers.  These are the constraints responsible for removing the potentially Boulware-Deser ghost-like modes.  As a theory with $\N$ interacting metrics will have a spectrum consisting of one massless graviton and $\N-1$ massive gravitons, there are  $\N-1$ potential Boulware-Deser ghosts, one for each massive graviton.  Here we see that for the interaction terms \eqref{V}, there are just the right number of extra primary constraints to vanquish all of them.  

Let us count degrees of freedom.  The phase space starts with $\N \times 2d^2$ degrees of freedom of the spatial vielbeins and their conjugate momenta. There is one overall spatial local Lorentz invariance generated by the first class constraints $\sum_I \PI_{ab}$, thus removing $2\times d(d-1)/2$ dimensions of phase space.  The other $\N-1$ combinations of Lorentz generators do not generate gauge symmetries and so will lead to secondary constraints, which together with the primary constraints will generate a second class set, removing an additional $2\times (\N-1)d(d-1)/2$ dimensions of phase space.  

There is one overall diffeomorphism, the generators of which enforce first class constraints, thus removing $2\times (d+1)$ dimensions of phase space.  Finally, we have the $\N-1$ additional primary constraints found above.  Though we will not show it here, we expect these constraints to each generate a secondary constraint, with which they should form a second class set, thus removing a further $2\times (\N-1)$ degrees of freedom.  Adding everything up, we find that the physical phase space should have
\be  2\left[\tfrac{1}{2}d(d-1)-1\right]+ 2(\N-1)  \left[\tfrac{1}{2}d(d+1)-1\right], \label{degoff}\ee
dimensions.  This corresponds exactly to one massless spin-2 field and $\N-1$ massive spin-2 fields and their canonical momenta, with no extra ghosts.

It should be emphasized that we have only shown the existence of the primary constraints necessary for eliminating the ghosts.  To complete the proof that these theories are ghost-free, the secondary constraints (which arise by demanding that the primary constraints be preserved in time) must be computed.  To show that there are not too few degrees of freedom, it must also be argued that there are no further constraints, and that the primary and secondary constraints together form a second class set.  However, it is hard to imagine that the secondary constraints would be absent, as they have been shown to exist in the bi-metric case (in the metric formalism) in \cite{Hassan:2011ea} and, what's more, their absence would indicate the existence of some kind of half degree of freedom which would not be consistent with Lorentz invariance in dimensions greater than 2.  (Though unlikely, it is possible that with multiple metrics, half degrees of freedom could pair up into full degrees of freedom.) 

The arguments here rely only on the fact that the Hamiltonian written in terms of the ADM variables is linear in all the lapse and shift variables, and goes through for the general interaction term \eqref{V}.  This includes any kind of theory graph, including tree graphs, loop graphs and graphs including tri-vertices, tetra-vertices, and beyond.  The argument is fully non-linear, and is valid to all orders in the fields and beyond any decoupling limits, such as those considered in \cite{deRham:2010ik}.

\section{Bi-gravity\label{bigravitysection}}
In this section we show that, for the case of $\N=2$ dynamical spin-2 fields, the vielbein theories introduced above are dynamically equivalent to the metric ghost-free bi-gravity theories, studied in \cite{Hassan:2011zd}.  The metric bi-gravity theory is obtained by promoting the flat reference metric $\eta_{\mu\nu}$ of dRGT massive gravity to a general metric $f_{\mu\nu}$ and allowing it to be dynamical by adding an Einstein-Hilbert term for $f_{\mu\nu}$, with its own Planck mass $M_{f}$,
\be  
\s = \intx \left[{M_{g}^{D-2}\over 2} \sqrt{-g}\,R[g]
+{M_f^{D-2}\over 2} \sqrt{-f}\,R[f]-{m^2M_{f\!g}^{D-2}\over 8}  \sqrt{-g}\,\sum_{n= 0}^D\beta_nS_n(\sqrt{g^{-1}f})\right]\, ,
\label{dRGTbimetricD}
\ee
where $M_{f\!g}^{D-2} \equiv (1/M_{g}^{D-2}+1/M_{f}^{D-2})^{-1}$.

There are $D+1$ different symmetric polynomials, and hence $D+1$ different parameters in the interaction term.  The first symmetric polynomial $\sqrt{-g} \, S_0(\sqrt{g^{-1}f}) = \sqrt{- g}$ and the $D$-th polynomial $\sqrt{-g} \, S_D(\sqrt{g^{-1}f}) = \sqrt{- f}$ are cosmological constants for $g$ and $f$, respectively.  Each polynomial gives a tadpole when expanded around flat space: $g_{\mu\nu} = \eta_{\mu\nu}+\tfrac{1}{M_g}\tilde g_{\mu\nu}$ and $f_{\mu\nu} = \eta_{\mu\nu}+\tfrac{1}{M_f}\tilde f_{\mu\nu}$.  If we demand that flat space is a solution for both metrics (i.e., no tadpoles), then we must take
\be
\sum_{k=0}^D \frac{\beta_k}{k!(D-k)!} =0\,, \ \ \ \ \ \sum_{k=1}^D \frac{\beta_k}{(k-1)!(D-k)!} =0\,.
\ee
For $D=4$ this means
\be
\beta_0 = -\(3\beta_1 +3\beta_2+\beta_3\)\, ,\ \ \ \ \  \beta_4 = - \( \beta_1+3\beta_2+3\beta_3 \) \, .
\ee
With this choice, expanding to quadratic order gives the Fierz-Pauli term \cite{Fierz:1939ix} for the fluctuation $\tfrac{1}{M_g}\tilde g_{\mu\nu}-\tfrac{1}{M_f}\tilde f_{\mu \nu}$.  The orthogonal fluctuation $\tfrac{1}{M_f}\tilde g_{\mu\nu}+\tfrac{1}{M_g}\tilde f_{\mu \nu}$ is absent from the potential.  Thus this theory propagates precisely one massive spin-2 field and one massless spin-2 field around flat space.  We can absorb one further coefficient by taking $m$ to be the mass of the massive spin-2 and setting
\be
(D-2)! \ \sum_{k=2}^D \frac{\beta_k}{(k-2)!(D-k)!} = -8\, .
\ee
For $D=4$, this gives
\be
\beta_1+2\beta_2+\beta_3= 8 \, .
\ee
In $D$ dimensions, the bi-gravity theory will have $D-2$ free parameters after tadpoles are eliminated and one coefficient is absorbed into the mass.

Despite the asymmetric appearance, the potential in \eqref{dRGTbimetricD} does not actually favor one metric over the other, since we have the property
\be 
\sqrt{-g}\,S_n(\sqrt{g^{-1}f})=\sqrt{-f}\,S_{D-n}(\sqrt{f^{-1}g})\,.
\label{dual}
\ee
If we wish, we can impose a ${\mathbb Z}_2$ symmetry under the interchange $g_{\mu\nu}\leftrightarrow f_{\mu\nu}$, by setting $M_{g}=M_{f}$ and $\beta_n=\beta_{D-n}$.

The two Einstein-Hilbert terms are each invariant under a separate diffeomorphism symmetry,
\be 
g_{\mu\nu}(x)\rightarrow {\partial \fone^\alpha \over \partial  x^\mu}{\partial  \fone^\beta \over \partial  x^\nu}g_{\alpha\beta}\(\fone(x)\) ,\ \ \  f_{\mu\nu}(x)\rightarrow {\partial \ftwo^\alpha \over \partial  x^\mu}{\partial \ftwo^\beta \over \partial  x^\nu}g_{\alpha\beta}\( \ftwo(x)\) \,.
\ee
The mass term breaks this down to the subgroup of diagonal diffeomorphisms with $\fone^\mu=\ftwo^\mu\equiv f^\mu$, which acts the same way on both metrics,
\be  
g_{\mu\nu}(x)\rightarrow {\partial f^\alpha \over \partial  x^\mu}{\partial  f^\beta \over \partial  x^\nu}g_{\alpha\beta}\( f(x)\) ,\ \ \  f_{\mu\nu}(x)\rightarrow {\partial {f}^\alpha \over \partial  x^\mu}{\partial  f^\beta \over \partial  x^\nu}g_{\alpha\beta}\( f(x)\) \,.
\ee
Naive counting arguments would suggest that a generic theory of two interacting spin-2 fields with only one diffeomorphism invariance will propagate 8 degrees of freedom for $D=4$: one massless spin-2, one massive spin-2 and one scalar ghost.  However, it was shown in \cite{Hassan:2011ea,Hassan:2011zd} that the theories \eqref{dRGTbimetricD} propagate only 7 degrees of freedom at the full non-linear level, consistent with one massless and one massive spin-2 alone.

\subsection{Vielbein formulation of bi-gravity}
We show now that the metric bi-gravity theory has an equivalent vielbein formulation.  We introduce two vielbeins, one for each metric,
\be
g_{\mu\nu} = \Eone_\mu^{\ A} \, \Eone_\nu^{\ B}\,\eta_{AB} \,  ,\ \ \  f_{\mu\nu} = \Etwo_\mu^{\ A} \, \Etwo_\nu^{\ B}\,\eta_{AB} \,  .
 \label{vielbimetric}
\ee
When the substitution \eqref{vielbimetric} is made into the Einstein-Hilbert part of the action, it becomes invariant under two separate local Lorentz transformations, 
\be 
\Eone_\mu^{\ A}\rightarrow \Lone^{A}_{\ B} \Eone^{\ B}_\mu,\ \ \ \Etwo_\mu^{\ A}\rightarrow \Ltwo^{A}_{\ B} \Etwo^{\ B}_\mu,
\ee
in addition to the two diffeomorphisms, under which the vielbeins transform as one-forms,
\be 
\Eone_{\mu}^{\ A}(x)\rightarrow {\partial \fone^\nu \over \partial  x^\mu}\Eone_{\nu}^{\ A}\left( \fone(x)\right) \, ,\ \ \ \Etwo_{\mu}^{\ A}(x)\rightarrow {\partial \ftwo^\nu \over \partial  x^\mu}\Etwo_{\nu}^{\ A}\left( \ftwo(x)\right) \, .
\ee

We will show that the following action, which consists of the Einstein-Hilbert terms in vielbein form and an interaction term written using wedge products of the vielbein one-forms $\Eb^A=E_\mu^{\ A}dx^\mu$, is dynamically equivalent to \eqref{dRGTbimetricD} in terms of the variables \eqref{vielbimetric},   
\bea 
\label{drgtvielbi}
\s&=& {M_g^{D-2}\over 2}\int d^Dx\ (\det\Eone) \,R[\Eone]+{M_f^{D-2}\over 2}\int d^Dx\ (\det\Etwo) \,R[\Etwo] \\ 
&& -{m^2M_{f\!g}^{D-2}\over 8}\int \sum_{n= 0}^D \frac{\beta_{n}}{n!(D-n)!} \tilde \epsilon_{A_1A_2\cdots A_D}\bEtwo^{ A_1}\wedge \cdots\wedge \bEtwo^{ A_n}\wedge \bEone^{ A_{n+1}}\wedge \cdots\wedge \bEone^{ A_D}\, . \nn
\eea
The mass term breaks the two diffeomorphism and local Lorentz symmetries down to the diagonal subgroup where ${\Lambda_{(1)}}^{A}_{\ B}={\Lambda_{(2)}}^{A}_{\ B}$ and ${f_{(1)}}^\mu={f_{(2)}}^\mu$.

Using the relation \eqref{dens}, the potential can be written in terms of the symmetric polynomials (defined in Appendix \ref{charpolyappendix2}),
\be
\sum_{n= 0}^D \beta_n  \(\det \Eone \) S_n(\Eone^{-1} \Etwo) \, .
\label{bipo}
\ee
Our first step is to show that the action with this potential, is in fact equivalent to the same action with the additional constraint that the following product of vielbeins and vielbein inverses be symmetric with respect to the Minkowski metric\footnote{Note that, taking the inverse of both sides of \eqref{symmvielbi}, we obtain $\Etwo^{-1}\Eone\, \eta= \eta\(\Etwo^{-1}\Eone\)^T$, so this condition is in fact symmetric under $1\leftrightarrow 2$.}.  In matrix notation,
\be
\label{symmvielbi}
\Eone^{-1}\Etwo\, \eta= \eta\(\Eone^{-1}\Etwo\)^T\,.
\ee
To see this equivalence, parametrize one of the vielbeins, say the first vielbein $\Eone$, as a generic Lorentz transformation times a constrained vielbein $\barEone$, chosen to satisfy \eqref{symmvielbi}, 
\be 
\label{vielbeinlparambi} 
\Eone=\barEone \, e^{\omega}\, , \ \ \ \ \ \barEone^{-1}\Etwo\, \eta= \eta\(\barEone^{-1}\Etwo\)^T\,.
\ee
We have written the Lorentz transformation as the exponential of a matrix $\omega$ which is anti-symmetric with respect to $\eta$,
\be
\eta\,\omega= -\omega^T\, \eta \, .
\ee
Equation \eqref{vielbeinlparambi} is nothing but a parametrization of the general vielbein -- we have packaged the $D^2$ components of the general vielbein into the $D(D+1)/2$ components of a constrained vielbein and $D(D-1)/2$ components of a Lorentz transformation.  

Now, we will see that the $D(D-1)/2$ variables in $\omega$ are auxiliary variables, and that their equations of motion set $\omega=0$.  Plugging the decomposition \eqref{vielbeinlparambi} into the action \eqref{drgtvielbi}, the parameters in the Lorentz transformation only appear through the potential, since the Einstein-Hilbert term is separately Lorentz invariant in each vielbein.   The potential takes the form
\be
\sum_{n= 0}^D \beta_n  (\det \barEone ) \, S_n\( e^{-\omega} \barEone^{-1} \Etwo\) \, .
\label{bipowomega}
\ee
Now vary with respect to $\omega$.  Consider first the lowest order terms in $\omega$ which contain no powers of $\omega$ beyond $\delta\omega$.
Due to the form of the symmetric polynomials, the only terms that appear at lowest order are traces of $\delta\omega$ with powers of the matrix $\barEone^{-1} \Etwo$.  Since $\barEone^{-1}\Etwo$ is symmetric and $\delta\omega$ antisymmetric (both with respect to $\eta$), and since $\delta\omega$ appears only linearly, the lowest order expressions in $\omega$ vanish.  Thus the equations of motion start linearly in $\omega$, and are solved by\footnote{There is also the possibility of having non-trivial solutions of the $\omega$ equations, in which case there would be more than one branch of the theory.  In this case,  the $\omega=0$ branch of the vielbein theory would be equivalent to the metric theory, and the other branches may not be.}
\be \omega=0\,.\ee
Thus the action with unconstrained vielbeins is dynamically equivalent to the action with constrained vielbeins.  We may plug $\omega=0$ into the potential \eqref{bipowomega},
\be
\sum_{n= 0}^D \beta_n   (\det \barEone) S_n(\barEone^{-1} \Etwo) \, .
\ee

Let us now relate the vielbein potential to the metric potential.  In matrix notation the metrics are given by $g = \Eone \, \eta \, \Eone^T$ and $f= \Etwo\, \eta \, \Etwo^T\,$.  Therefore, 
\be
g^{-1} f= (\Eone^{-1})^T \, \eta^{-1} \,\Eone^{-1}\,\Etwo\, \eta \, \Etwo^T\,.
\ee
Using the parametrization \eqref{vielbeinlparambi}, along with the symmetry property of $\barEone$ gives $g^{-1} f=\barEone^{-1T}\Etwo^T \barEone^{-1T}\Etwo^T$, or
\be
\sqrt{g^{-1} f} =\(\Etwo \,  \barEone^{-1} \)^T\,.
\ee  
Using the properties \eqref{transposeprop} and \eqref{flipprop}, we may write
\be
 (\det \barEone)\, S_n(\barEone^{-1} \Etwo)=
 \sqrt{-\det g}\,\, S_n(\sqrt{g^{-1} f }) \,.
\ee
We see that the vielbein bi-gravity theory is dynamically equivalent to the metric bi-gravity theory, as claimed.

Note that we have only proved the equivalence of the metric and vielbein formulation for theories with one bi-gravity interaction.  When constructing multi-gravity theory graphs out of the bi-vertex as in figure \ref{generalgraph}, the equivalence of the vielbein formulation and the metric formulation does not appear to always hold.  In Appendix \ref{Lorentz} we argue that the equivalence holds as long as the theory graph is a tree graph, i.e., contains no closed loops.  If the theory graph contains loops however, the equivalence appears to break down.  This is directly related to the form of the constraints used to pass from the vielbein formulation to the metric formulation \eqref{symmvielbi}.  Essentially, the loop graphs alter the form of these constraints.

In the vielbein formulation, the proof of ghost-freedom is trivial to extend to all theory graphs.  Thus we expect graphs such as figure \ref{generalgraph} to be ghost-free when the interactions are given in the vielbein formulation.  For the metric theory however, it is not known whether the proof of ghost-freedom extends to the case of theory graphs with closed loops.  The multi-gravity theory represented in figure \ref{generalgraph}, or, for example, a ``triangle" theory such as that studied in \cite{Khosravi:2011zi}, potentially contains ghosts when formulated in terms of metrics rather than vielbeins.  (See Appendix \ref{Lorentz} for more on this point.)

 \section{dRGT massive gravity}
We now turn to the case of massive gravity.  We treat this case separately not only because of specific interest in massive gravity but also because these theories do not inherently possess the overall diffeomorphism invariance of the multi-metric theories.  Thus in this section we will repeat the Hamiltonian analysis specifically for massive gravity.  We will see that demonstrating the existence of the primary constraint which eliminates the Boulware-Deser ghost is much easier in vielbein variables.

In $D$ spacetime dimensions the action for ghost-free dRGT massive gravity is 
\be 
\s = {M_P^{D-2}\over 2}\int d^Dx\ \sqrt{-g}\left[R-{m^2\over 4} \sum_{n= 0}^D\beta_nS_n(\sqrt{g^{-1}\eta})\right]\,.
\label{dRGTtheoryD}
\ee
The massive gravity theory depends on the dynamical metric $g_{\mu\nu}$ and the fixed background metric $\eta_{\mu\nu}$.  The Einstein-Hilbert part of the action \eqref{dRGTtheoryD} is invariant under diffeomorphisms $f^\mu(x)$, 
\be
\label{GRfullgauge1}
 g_{\mu\nu}(x)\rightarrow {\partial f^\alpha \over \partial  x^\mu}{\partial f^\beta \over \partial  x^\nu}g_{\alpha\beta}\(f(x)\) \,,
 \ee
but the mass term breaks this symmetry, due to the appearance of the background metric $\eta_{\mu\nu}$.  

As in the bi-metric case, there are $D+1$ different symmetric polynomials, and hence $D+1$ different parameters in the mass term.  For massive gravity, the $D$-th symmetric polynomial is just $\sqrt{-\det \eta}=1$, so it doesn't contribute to the equations of motion.  Thus there are only $D$ free parameters.  We can ensure that flat space is a valid solution by demanding
\be
D! \ \sum_{k=0}^D \frac{\beta_k}{k!(D-k)!} = (D-1)! \ \sum_{k=1}^D \frac{\beta_k}{(k-1)!(D-k)!} \,.
\ee
For $D=4$, this gives
\be
\beta_0 = -\(3\beta_1 +3\beta_2+\beta_3\)\, .
\ee
Then, expanding to quadratic order gives the Fierz-Pauli term \cite{Fierz:1939ix} for the fluctuation $h_{\mu\nu}=g_{\mu\nu}-\eta_{\mu\nu}$.  Thus this theory propagates precisely one massive spin-2 field around flat space.  We again absorb one further coefficient by taking $m$ to be the mass of the massive spin-2 and setting
\be
-D! \ \sum_{k=0}^D \frac{\beta_k}{k!(D-k)!} +(D-2)! \ \sum_{k=2}^D \frac{\beta_k}{(k-2)!(D-k)!} = -8\, .
\ee
For $D=4$, this gives
\be
\beta_1+2\beta_2+\beta_3 = 8 \, .
\ee
In $D$ dimensions, the theory has $D-2$ free parameters in addition to the mass.

\subsection{Vielbein formulation of massive gravity}
Our goal now is to show that dRGT massive gravity in $D$ dimensions \eqref{dRGTtheoryD} has a dynamically equivalent vielbein formulation\footnote{Vielbein massive gravity is also considered in \cite{Nibbelink:2006sz}.}.  We introduce the vielbein fields
$E_\mu^{\ A}(x)$,
\be
\label{fielbeindef}
g_{\mu\nu} = E_\mu^{\ A} \, E_\nu^{\ B}\,\eta_{AB} \,  , 
\ee
along with the following action, which consists of the Einstein-Hilbert term in vielbein form, and potential terms written using wedge products of the vielbein one-forms $\Eb^A=E_\mu^{\ A}dx^\mu$, and unit one-forms which can be thought of as vielbeins for the flat background metric ${\bf 1}^A=\delta_\mu^{\ A} dx^\mu$,
\begin{multline}  
\s = {M_P^{D-2}\over 2}\Bigg(\int d^Dx\ \det(E) R[E]\\
-{m^2\over 4}\int \sum_{n= 0}^D \frac{\beta_{n}}{n!(D-n)!} \tilde \epsilon_{A_1A_2\cdots A_D} {\bf 1}^{ A_1}\wedge \ldots\wedge  {\bf 1}^{ A_n}\wedge \Eb^{ A_{n+1}}\wedge \ldots\wedge \Eb^{ A_D}\Bigg) \, . 
\label{drgtviel}
\end{multline}
By pulling out a determinant as in \eqref{dens}, the mass term can also be written in terms of the symmetric polynomials of the matrix $E_\mu^{\ A}$,
\be
\frac{1}{n!(D-n)!} \tilde\epsilon_{A_1A_2\cdots A_D} {\bf 1}^{ A_1}\wedge \ldots\wedge  {\bf 1}^{ A_n}\wedge\Eb^{ A_{n+1}}\wedge \ldots\wedge \Eb^{ A_D} =(\det E\,d^Dx) S_{n}(E^{-1}) \, .
 \label{charpolyveilbe}
\ee

The action \eqref{drgtviel} is an action for $D^2$ variables, whereas \eqref{dRGTtheoryD} is an action for $D(D+1)/2$ variables.  Furthermore, \eqref{drgtviel} has no gauge symmetry, since both diffeomorphisms and LLI are broken by the mass term.  Nevertheless, we will show that \eqref{drgtviel} is dynamically equivalent to \eqref{dRGTtheoryD}.  The logic proceeds much as it did for the bi-gravity case.

We show that the action \eqref{drgtviel} is dynamically equivalent to the same action with the additional constraint that the vielbein be symmetric with respect to the Minkowski metric,
\be
\label{symmviel}
E\, \eta= \eta\, E^T\,.
\ee
We parametrize the general vielbein as a constrained vielbein $\barE$ satisfying \eqref{symmviel}, times a generic Lorentz transformation\footnote{See \cite{Deffayet:2012zc} for more discussion on this condition and its relation to the matrix square roots of the metric formulation.}, 
\be 
\label{vielbeinlparam} 
E=\barE \, e^{\omega}\,.
\ee
Again, the Lorentz transformation is written as the exponential of a matrix $\omega$ which is anti-symmetric with respect to $\eta$, 
\be \eta\,\omega= -\omega^T\, \eta.\ee
The $\omega$'s will not appear through the Einstein-Hilbert term, since it is invariant under local Lorentz transformations.

As in bi-gravity, the $D(D-1)/2$ variables in $\omega$ appear only through the mass term and are auxiliary variables.  Their equations of motion will set $\omega=0$.  Plugging the decomposition \eqref{vielbeinlparam} into the action \eqref{drgtviel}, the parameters in the Lorentz transformation appear through the mass term,
\be
\sum_{n= 0}^D \beta_n  (\det \barE ) \, S_n\( e^{-\omega} \barE^{-1} \) \, .
\label{powomega}
\ee
Now consider the $\omega$ equations of motion.  Vary with respect to $\omega$ and consider the equations of motion in powers of $\omega$.  The lowest order terms  contain no powers of $\omega$ beyond the variation $\delta\omega$. The only terms that appear at lowest order in the symmetric polynomials are traces of $\delta\omega$ with powers of $\barE^{-1}$.  Since $\barE^{-1}$ is symmetric and $\delta\omega$ antisymmetric (both with respect to $\eta$), and since $\delta\omega$ appears only linearly, the terms in the equations of motion linear in $\omega$ vanish.  Thus the equations of motion start linearly in $\omega$, and are solved by\footnote{As in the bi-metric case, there is the possibility of additional branches if there are non-trivial solutions of the $\omega$ equations.  If these exist, then the equivalence between vielbein and metric theories may only hold for the trivial branch.}
\be \omega=0.\label{omegazerosolm}\ee
We may plug the solution \eqref{omegazerosolm} into the potential \eqref{powomega}, giving 
\be \label{veilbeinpotcons}
\sum_{n= 0}^D \beta_n   (\det \barE) S_n(\barE^{-1}) \, .
\ee
Thus the action with unconstrained vielbeins is dynamically equivalent to the action with vielbeins constrained to satisfy \eqref{symmviel}.  

To relate the potential \eqref{veilbeinpotcons} to the metric potential we write $g = E\, \eta \, E^T$ so that
\be
g^{-1} \eta= (E^{-1})^T \, \eta^{-1} \,E^{-1}\,\eta\,.
\ee
Using the parametrization \eqref{vielbeinlparam} and the symmetry property of $\barE^{-1}$, we see that
\be
\sqrt{g^{-1} \eta} =\( \barE^{-1} \)^T\,.
\ee  
Thus we can write, using the property \eqref{transposeprop} of the symmetric polynomials,
\be
 (\det \barE)\, S_n(\barE^{-1})=
 \sqrt{-\det g}\,\, S_n(\sqrt{g^{-1} \eta}) \,.
\ee
We see that the vielbein massive gravity is equivalent to dRGT massive gravity.

\subsection{Ghost-freedom\label{dRGTghostsec}}

We've seen that we can write the action for dRGT massive gravity using the $D^2$ components of an unconstrained vielbein as variables.  Now, by choosing a different parametrization for the vielbein, we will see that it is almost trivial to identify the primary constraint which eliminates the Boulware-Deser ghost.  

We again perform a $d+1$ decomposition of the general vielbein as in \eqref{uptrianparam}, an upper triangular vielbein rotated by a standard Lorentz boost parametrized by $p^a$,
 \be
E_\mu^{\ A} =
\left(\begin{array}{cc}N \gamma +N^i e_i^{\ a} p_a& N p^a+N^i e_i^{\ b} (\delta_b^{\ a}+\frac{1}{\gamma+1}p_b p^a) \\
e_i^{\ a} p_a &  e_i^{\ b} (\delta_b^{\ a}+\frac{1}{\gamma+1}p_b p^a)\end{array}\right) \,.
 \ee 
The $p^a$ do not enter the Einstein-Hilbert term, since it is Lorentz invariant.

The mass term \eqref{charpolyveilbe} is not invariant under local Lorentz transformations, so there will be explicit dependence on the $p^a$'s.  Note, however, that the lapse and shift, $N$ and $N^i$, only appear in the components $E_0^{\ 0}$ and $E_0^{\ b}$, and that they both appear linearly.  Due to the epsilon tensor in the mass term, there will never be more than one component $E_0^{\ 0}$ or $E_0^{\ b}$ in any term, and so the entire interaction term is manifestly linear in both the lapse and the shift $N$ and $N^i$.  Thus we can write the mass term in the form
\be 
U=N{\cal C}^\m(\de,p)+N^i{\cal C}^\m_i(\de,p)+{\cal H}(\de,p) \, .
\ee
As the Einstein-Hilbert Hamiltonian is also linear in the lapse and shift, these remain Lagrange multipliers in the full massive theory, enforcing the constraints
\be 
{\cal C}(\de,\pi)+{\cal C}^\m(\de,p)=0,\ \ \ {\cal C}_i(\de,\pi)+{\cal C}_i^\m(\de,p)=0 \,.
\ee
Note that in the metric formulation, the lapse and shift do not automatically appear in this way.

The $p^a$'s are auxiliary variables, since they appear appear only through the mass term, with no derivatives.  We can eliminate them by solving the $N^i$ constraints, 
\be 
{\cal C}_i(\de,\pi)+{\cal C}_i^\m(\de,p)=0 \Rightarrow p^a=p^a(\de,\pi)\,.
\ee
The action now takes the form
\be 
\s =\intx \left(  \pi^i_{\ a}\dot e_i^{\ a}-{\cal H}(\de,p(\de,\pi))-{1\over 2}\lambda^{ab}{\cal P}_{ab}(\de,\pi)-N\left[{\cal C}(\de,\pi)+{\cal C}^\m(\de,p(\de,\pi))\right]\right)\, . 
\ee
We see explicitly the presence of the extra primary constraint, enforced by the shift $N$, responsible for removing the Boulware-Deser ghost.  As shown in \cite{Hassan:2011ea}, there will be an additional secondary constraint which arises by demanding that the primary constraint enforced by $N$ be conserved in time. 

Our theory has no gauge symmetry -- no diffeomorphism symmetry or local Lorentz invariance -- so there will also be secondary constraints which come from demanding that the primary Lorentz constraints ${\cal P}_{ab}$ be preserved in time. These combine with the primary constraints into a second class set.   

Let us now count degrees of freedom:  the phase space has $2d^2$ variables, the $d^2$ spatial components of the vielbein $e_i^{\ a}$, and their $d^2$ canonical momenta $\pi^i_{\ a}$.  There are then second class constraints restricting the phase space: $2\times {d(d-1)\over 2}$ constraints coming from the local Lorentz constraints and their secondary constraints, and $2\times 1$ special constraints coming from the lapse multiplier and the corresponding secondary constraint.  This leaves a 
\be 2\left({d(d+1)\over 2}-1\right)\label{massivedimphase}\ee
dimensional physical phase space, just right to describe the degrees of freedom of a traceless symmetric tensor and its canonical momenta, i.e. a massive spin-2 graviton, with no extra Boulware-Deser modes.

Given a background solution to the bi-gravity theory (\ref{dRGTbimetricD}) where one of the metrics is Minkowski, we can recover the massive gravity theory (\ref{dRGTtheoryD}) by sending the Planck mass associated with the Minkowski metric to infinity, after canonically normalizing the fluctuations.  This can be done for any background other than Minkowski, and the result will be ghost-free massive gravity propagating on that background.  In fact, many different theories containing only massive gravitons, around various and even multiple backgrounds, can be obtained by taking the appropriate $\MI\rightarrow \infty$ limits of the general multi-metric theory.  By arguments similar to those in this section, it can be seen that all these theories are ghost-free.

\section{Discussion}

In this paper, we have reformulated the recently uncovered theories of massive gravity and bi-gravity in terms of vielbein variables.  We find that with this choice of variables, the theories become much more natural and transparent.  In place of the unwieldy polynomials and matrix square roots of the metric formulation, the vielbein formulation simply contains wedge products of the possible combinations of the vielbeins.  In vielbein form, the additional primary constraints which signal the absence of the Boulware-Deser ghost are manifest, to all orders in the fields and beyond any decoupling limits.

The vielbein formulation has allowed us to extend the ghost-free interactions to interactions between multiple metrics.  The natural extension is to allow for arbitrary wedge products between all possible vielbeins, leading to the general interaction \eqref{V}.  We have seen that the proof of the existence of additional primary constraints extends easily to this general interaction.

We conjecture that the interaction \eqref{V} is in fact the most general ghost-free zero-derivative interaction among multiple vielbeins, and thus the most general such interaction among massive and massless spin-2 fields.  This interaction can be though of as a kind of multi-metric generalized cosmological constant, or a lowest-order generalized multi-Lovelock invariant\footnote{Note that the addition of the usual single-metric Lovelock invariants to the multi-gravity theories does not disrupt the proof of ghost-freedom \cite{Paulos:2012xe}.} \cite{Lovelock:1971yv,Lovelock:1972vz}.  We will expound on this relation in future work.

This construction opens up a wide parameter space in which to search for phenomenologically interesting or promising models, and in which to search for theories which avoid potential problems, such as superluminality around non-trivial solutions, and low scale strong coupling.  There are parallels in galileon theory: as first shown in \cite{deRham:2010ik,deRham:2010kj} the decoupling limit of dRGT gives the single field galileon theories of \cite{Nicolis:2008in}.  Multi-metric theories should yield some kind of multi-galileon along the lines of \cite{Deffayet:2010zh,Padilla:2010de,Hinterbichler:2010xn} (though many of these multi-galileon theories also have instabilities or superluminality \cite{Padilla:2010tj,Andrews:2010km,Fromont:2013fk,Garcia-Saenz:2013vn}).  It has been argued that for a single massive graviton of mass $m$ in $D=4$, the highest unitarity bound which is possible is the rather low cutoff $\Lambda_3\equiv \(M_P m^2\)^{1/3}$ \cite{Schwartz:2003vj}.  It is possible that within these interacting multi-metric theories, there are examples with higher cutoffs.   

Finally, we comment on the possible coupling of multiple metrics to matter.  If matter is minimally coupled to only a single metric, then clocks and rulers will measure distances as determined by that single metric.  This kind of coupling will not re-introduce the Boulware-Deser ghost, as it maintains the same symmetries of the Einstein-Hilbert term.  However, more general ghost-free couplings might exist that involve more than one metric and maintain an overall diffeomorphism invariance (we might call this multi-minimal coupling), though this can naively be expected to violate the equivalence principle.  

\vskip.5cm

\bigskip
{\bf Acknowledgements}: 
We would like to thank F. Hassan and A. Schmidt-May for helpful discussions and G. Gabadadze for comments of the draft.   R.A.R. would like to thank the Center for Particle Cosmology at the University of Pennsylvania for its hospitality, during which time this work was initiated.  Both authors would like to thank the organizers of the conference ``Essential Cosmology for the Next Generation,'' as well as Princeton University and 15 Hamilton for hospitality during the progression of this work.  R.A.R. is supported by NASA under contract NNX10AH14G and by the US Department of Energy under grant DE-FG02-11ER41743.  K.H. is supported by NASA ATP Grant NNX08AH27G, and by funds provided by the University of Pennsylvania.

\appendix

\section{Symmetric polynomials\label{charpolyappendix}}

In this appendix, we define the various matrix polynomials used throughout the paper, and describe some of their properties. 

\subsection{Elementary symmetric polynomials\label{charpolyappendix2}}

Given a $D\times D$ matrix $M^A_{\ B}$, we define the elementary symmetric polynomials, for $0\leq n\leq D$, 
\be
\label{e}
S_n(M) = \frac{1}{n ! (D-n)!}\, 
\tilde\epsilon_{A_1 A_2\cdots A_D}\, 
\tilde\epsilon^{B_1B_2 \cdots B_D}\, 
M^{ A_1}_{\ B_1}\cdots M^{ A_n}_{\ B_n}
\delta^{ A_{n+1}}_{\ B_{n+1}}\cdots \delta^{ A_D}_{\ B_D}\,,
\ee
or equivalently
\be  \label{symepsexp}
M^{ A_1}_{\ B_1}\cdots M^{ A_n}_{\ B_n} \delta^{ A_{n+1}}_{\ B_{n+1}}\cdots \delta^{ A_D}_{\ B_D}\tilde\epsilon_{A_1 A_2\cdots A_D}={n! (D-n)!\over D!} S_n(M) \tilde\epsilon_{B_1B_2 \cdots B_D}.
\ee

In terms of traces of products of $M$, the first few are
\bea S_0(M)&=&1 \, ,\nn\\
S_1(M)&=&[M] \, ,\nn\\
S_2(M)&=&{1\over 2!}\left( [M]^2-[M^2]\right)\, ,\nn\\
S_3(M)&=&{1\over 3!}\left( [M]^3-3[M][M^2]+2[M^3]\right) \, ,\nn\\
S_4(M)&=&{1\over 4!}\left( [M]^4-6[M]^2[M^2]+8[M][M^3]+3[M^2]^2-6[M^4]\right)\, , \nn\\
&&\vdots
\eea
The $D$-th symmetric polynomial is the determinant,
\be S_{D}(M) =\det M \, ,\ee
and the higher symmetric polynomials are defined to vanish identically,
\be S_n(M)= 0 ~~ {\rm for}~~ n>D \, .\ee

If $M$ is diagonalizable, the symmetric polynomials are the symmetric polynomials in the eigenvalues.  If we label the eigenvalues (including degeneracy) $\lambda_A$, $A=1,\cdots,D$, then
\bea S_0(M)&=&1 \, ,\nn\\
S_1(M)&=&\sum_A \lambda_A \, ,\nn\\
S_2(M)&=&\sum_{A < B}\lambda_A\lambda_B\, ,\nn\\
S_3(M)&=&\sum_{A < B<C}\lambda_A\lambda_B\lambda_C\, ,\nn\\
&& \vdots \nn\\
S_D(M)&=&\lambda_1\lambda_2\cdots\lambda_D.
\eea

The symmetric polynomials may be obtained from expanding out the following determinant, in powers of $\epsilon$,
\be \det\(1+\epsilon M\)=\sum_{n=0}^D\epsilon^n S_n(M).\ee

Transposing a matrix leaves its symmetric polynomial unchanged,
\be S_n(M^T)=S_n(M),\label{transposeprop}\ee
and given the symmetric polynomial of a product of two $D\times D$ matrices $M$ and $N$, we may cyclically permute the argument,
\be S_n(MN)=S_n(NM).\label{flipprop}\ee

\subsection{Generalized symmetric polynomials\label{charpolyappendix3}}

Define the generalized symmetric polynomials for multiple $D\times D$ matrices $M^{(1)}, M^{(2)},\cdots, M^{(D)}$, as follows,
\be S\(M^{(1)},\ M^{(2)},\cdots, M^{(D)}\)=\tilde\epsilon_{A_1 A_2\cdots A_D}\, 
\tilde\epsilon^{B_1B_2 \cdots B_D}\, 
{M^{(1)}}^{ A_1}_{\ B_1} {M^{(2)}}^{ A_2}_{\ B_2}\cdots {M^{(D)}}^{ A_D}_{\ B_D}\,,
\label{multichardef}
\ee
or equivalently
\be 
{M^{(1)}}^{ A_1}_{\ B_1} {M^{(2)}}^{ A_2}_{\ B_2}\cdots {M^{(D)}}^{ A_D}_{\ B_D}\tilde\epsilon_{A_1 A_2\cdots A_D}={1\over D!}S\(M^{(1)},\ M^{(2)},\cdots, M^{(D)}\)\tilde\epsilon_{B_1B_2 \cdots B_D}\, .
\ee
We can calculate $S\(M^{(1)},\ M^{(2)},\cdots, M^{(D)}\)$ by writing all $D!$ of the top to bottom contractions of the indices of the $M$'s, with a sign for straightening out the contraction.  For example:
\begin{itemize}
\item for $D=2$, there are two possible contractions, and we have
\be S\(\mX,\mY\)=[\mX][\mY]-[\mX \mY],\ee
\item
for $D=3$ there are six possible contractions, and we have
\be  S\(\mX,\mY,\mZ\)=[\mX][\mY][\mZ]-[\mX][\mY\mZ]-[\mY][\mZ\mX]-[\mZ][\mX\mY]+[\mX\mY\mZ]+[\mX\mZ\mY],\ee
\item
for $D=4$ there are 24 possible contractions 
\bea
\label{ABCD}
&&\!\!\!\! S\(\mX,\mY,\mZ,\mW\)=[\mX][\mY][\mZ][\mW]  - [\mX\mY\mZ\mW] - [\mX\mY\mW\mZ] -[\mX\mZ\mY\mW]
 \nonumber \\
&& \!\!\!\!  - [\mX\mZ\mW\mY] - [\mX\mW\mY\mZ] -[\mX\mW\mZ\mY] 
+[\mX] \Big([\mY\mZ\mW]+[\mY\mW\mZ]\Big)
+[\mY] \Big([\mX\mZ\mW]+[\mX\mW\mZ]\Big)  
 \nonumber \\
&&\!\!\!\! +[\mZ] \Big([\mX\mY\mW]+[\mX\mW\mY]\Big)
+[\mW] \Big([\mX\mY\mZ]+[\mX\mZ\mY]\Big)   -[\mX][\mY][\mZ\mW]-[\mX][\mZ][\mY\mW]\nonumber \\
&&\!\!\!\! -[\mX][\mW][\mY\mZ]
-[\mY][\mZ][\mX\mW]  -[\mY][\mW][\mX\mZ]-[\mZ][\mW][\mX\mY]\nonumber\\
&&+[\mX\mY][\mZ\mW]+[\mX\mZ][\mY\mW]+[\mX\mW][\mY\mZ] \, . \nn\\
\eea
\end{itemize}

The elementary symmetric polynomials defined in \eqref{e} are the special case,
\be S_n(M)=\frac{1}{n!(D-n)!} S(\underset{n\ {\rm times}}{\underbrace{M,\cdots,M}},\underset{D-n\ {\rm times}}{\underbrace{1,\cdots,1}}).\ee

\section{Hamiltonian formulation of GR in vielbein variables\label{GRveilbeinham}}

Here we review the Hamiltonian formulation of GR, in terms of upper triangular vielbein variables \eqref{uppertrianvielbein} (see also \cite{Deser:1976ay} and section 2.3 of \cite{Peldan:1993hi}), 
\be 
\label{uppertrianvielbein2}
\hat{E}_\mu^{\ A}=\left(\begin{array}{cc}N & N^ie_i^{\ a} \\0 & e_i^{\ a}\end{array}\right),\ \ \  \hat{E}^\mu_{\ A}=\left(\begin{array}{cc}{1\over N} & 0 \\-{N^i\over N} & e^i_{\ a}\end{array}\right)\,.
\ee
The Lagrangian form of the action is
\be\label{lagactionap}
\s={M_P^{D-2}\over 2}\intx (\det E) \ R(E)
={M_P^{D-2}\over 2}\intx \det(\de) \,N \left(R[\de]-K^{2}+K^{ij}K_{ij}\right) \, ,
\ee
where surface terms are omitted and the extrinsic curvature is written in terms of vielbeins, 
\be
K_{ij}={1\over 2N}\left(\dot e_i^{\ a}e_{ja}+\dot e_j^{\ a}e_{ia}-\nabla_i N_j-\nabla_j N_i\right)\,.
\ee
The covariant derivatives are with respect to the spatial metric $g_{ij}$, expressed in terms of the spatial vielbeins through $g_{ij}=e_i^{\ a}e_j^{\ b}\delta_{ab}$.

Now, we Legendre transform with respect to the spatial components of the vielbein, $e_i^{\ a}$.  The canonical momenta are
\be
\pi^i_{\ a}={\delta {\cal L}\over \delta \dot e_i^{\ a}}=M_P^{D-2}\,\det(\de)e_{ja}\left(K^{ij}-Kg^{ij}\right) \, .
\ee
Multiplying this by an $e_{ib}$ and realizing that $K_{ij}$ is symmetric, we arrive at a set of primary constraints,
\be
{\cal P}_{ab}=e_{i[a}\pi^i_{\ b]}=0 \,.
\label{rot}
\ee
The ${\cal P}_{ab}$ are anti-symmetric and represent $\tfrac{1}{2}d(d-1)$ constraints.

We can invert for $K_{ij}$ in terms of $\pi^i_{\ a}$,
\be
\label{kprelationf}
K_{i}^{\ j}={1\over M_P^{D-2} \det(\de)}\left(e_i^{\ a}\pi^{j}_{\ a}-{1\over D-2}\pi^k_{\ b}e_k^{\ b} \delta_{i}^j\right)\,.
\ee
From the definition of $K_{ij}$, we have 
\be
\label{edotdet}
\dot e_{(i}^{\ \ a}e_{j)a}=NK_{ij}+\nabla_{(i}N_{j)} \,.
\ee

Now we can calculate the Hamiltonian density,
\bea
{\cal H}&&=\pi^i_{\ a}\dot e_i^{\ a}-{\cal L}\\
&&={M_P^{D-2}\over 2}\det(\de)\big[2 e_{ja}\dot e_i^{\ a}\left(K^{ij}-Kg^{ij}\right) -R[\de]+K^{2}-K^{ij}K_{ij}\big] \, .
\eea
Using (\ref{edotdet}) and the symmetry of $K_{ij}$ to express the first term in terms of $K$'s, 
\be
2 e_{ja}\dot e_i^{\ a}\left(K^{ij}-Kg^{ij}\right)=\left(K^{ij}-Kh^{ij}\right)(2NK_{ij}+2\nabla_{i}N_{j}) \, ,
\ee
we find, after integrating by parts the terms containing $N_i$,
\be
{\cal H}={M_P^{D-2}\over 2}\left[N \det(\de)\left(-R[\de]-K^{2}+K^{ij}K_{ij}\right)-2N_i\det(\de)\nabla_j\left(K^{ij}-Kg^{ij}\right)\right]\, .
\label{H}
\ee
Now, using the expression (\ref{kprelationf}), we have expressed the Hamiltonian in terms of the canonical momenta and vielbein.

Adding Lagrange multipliers $\lambda^{ab}$ for the primary constraints \eqref{rot}, the GR action is now in the form of a constrained Hamiltonian system,
\bea
\s=\intx \(\pi^i_{\ a}\dot e_i^{\ a}-N{\cal C}-N_i{\cal C}^i-{1\over 2}\lambda^{ab}{\cal P}_{ab}\)\, ,
\eea
where
\be
{\cal C}={M_P^{D-2}\over 2}\det(\de)\left(R[\de]+K^{2}-K^{ij}K_{ij}\right),\ \ \ {\cal C}^i={M_P^{D-2}\over 2}\, 2\det(\de)\nabla_j\left(K^{ij}-Kg^{ij}\right),\ \ \  {\cal P}_{ab}=e_{i[a}\pi^i_{\ b]} \, .
\ee
Here's how the counting of constraints works in pure GR: the phase space is $2d^2$ dimensional, containing the $d^2$ components of the spatial vielbein and their canonical momenta.   The lapse and shift appear as Lagrange multipliers, enforcing the $d$ constraints ${\cal C}^i=0$ and the 1 constraint ${\cal C}=0$.  These are the diffeomorphism constraints associated to spatial reparametrizations and time reparametrizations, respectively.  On top of that, we have the $d(d-1)/2$ additional primary constraints ${\cal P}_{ab}$, which generate the spatial local Lorentz symmetries of the upper triangular vielbein.
These constraints are all first class, so each removes two dimensions from the phase space.  We are left with a $2\({(d-1)d\over 2}-1\)$ dimensional phase space, just right for describing the degrees of freedom of a transverse symmetric traceless tensor mode, i.e. a massless graviton.

\section{Lorentz constraints and vielbein/metric equivalence \label{Lorentz}}
A theory of $\N$ interacting spin-2 fields should maintain the same number of degrees of freedom when written in terms of vielbeins or in terms of metrics.  A metric contains $\frac{1}{2} D(D+1)$ components while a vielbein contain $D^2$ components, giving an extra $\frac{1}{2}D(D-1)$ components.  The interaction terms containing $\N$ vielbeins considered in this paper have only one local Lorentz invariance, i.e. the overall invariance which rotates all the vielbeins together.  This only accounts for the removal of the $\frac{1}{2}D(D-1)$ of one of the vielbeins.   The remaining $\N-1$ vielbeins must have their extra components removed in a different manner, and we will see in this Appendix how this comes about at the level of the Lagrangian.  

The extra components of the vielbein are removed by on-shell Lorentz constraints.  To see this, let us consider first the usual Einstein-Hilbert kinetic term written in terms of the vielbein,
\be
\s_{EH}={1\over 2}\intx \(\det E\) \,  R[E] \, .
\ee
This term is invariant under the local Lorentz transformation
\be
E_\mu^{~A} \rightarrow {E'}_\mu^{~A} = \Lambda_{\ B}^{A}E_\mu^{~B} \, .
\ee
Denote the infinitesimal Lorentz transformation by
\be
\Lambda_{\ B}^{A} \simeq \delta_B^{A} + \omega_{\ B}^{A} \, ,
\ee
where $\omega_{\ B}^{A}$ satisfies
\be
\eta^{CB}\, \omega_{\ B}^{A} = - \eta^{AB}\, \omega_{\ B}^{C}  \, \, .
\ee
Since the kinetic term is invariant under the local Lorentz transformation, the variation of the action under this transformation is zero,
\be
\delta \s_{EH}=
{1\over 2} \intx\,\frac{\delta (\det E\, R[E])}{\delta E_\mu^{~A}}\, \delta E_\mu^{~A} 
 =0\, ,
\ee
when $\delta E_\mu^{~A} = \omega_{\ B}^{A}E_\mu^{~B}$.

Since $\omega$ is an arbitrary anti-symmetric function of the spacetime coordinates, this implies that the following is identically zero off-shell
\be\label{EHeinstensymm}
 \frac{\delta (\det E\, R[E])}{\delta E_\mu^{~ A}}
E_\mu^{~C} \, \eta_{CB}-\(A\leftrightarrow B\)=0 \, .
\ee

Consider now the multi-vielbein theory, $\Eone, \Etwo, \ldots$, with an Einstein-Hilbert kinetic term for each vielbein, as well as a potential term $U(\Eone, \Etwo, \ldots)$ that mixes the $\EI$ but leaves precisely one overall Lorentz invariance intact.  The $\EI$ equations of motion for this theory are
\be
{1\over 2}\frac{\delta (\det \EI\, R[\EI])}{\delta \EI_\mu^{~A}} - \frac{\delta U(\Eone,\Etwo,  \ldots)}{\delta \EI_\mu^{~A}} = 0 \, .
\ee
Multiplying both terms by $\EI_\mu^{~C} \, \eta_{CB}$ and anti-symmetrizing, we find that the Einstein-Hilbert part vanishes due to \eqref{EHeinstensymm}, and the equations of motion imply
\be
\frac{\delta U(\Eone,\Etwo,  \ldots)}{\delta \EI_\mu^{~A}} \EI_\mu^{~C} \, \eta_{CB}-\(A\leftrightarrow B\) =0\, .
\ee
 
For each vielbein $\EI$, we therefore have one on-shell constraint, saying that the derivative of the potential times the vielbein is symmetric,
\be
\label{c}
\frac{\delta U(\Eone,\Etwo,  \ldots)}{\delta \EI_\mu^{~A}} \EI_\mu^{~C} \, \eta_{CB}= 
\frac{\delta U(\Eone,\Etwo,  \ldots)}{\delta \EI_\mu^{~B}} \EI_\mu^{~C} \, \eta_{CA} \, .
\ee
These constraints eliminate $\tfrac{1}{2}D(D-1)$ components of each vielbein.  These constraints are invariant under the overall local Lorentz invariance.

Note, however, that the sum of the constraints is identically satisfied, due to the one overall Lorentz invariance of the potential
\be
\sum_I \, \frac{\delta U(\Eone,\Etwo,  \ldots)}{\delta \EI_\mu^{~A}} \EI_\mu^{~C} \, \eta_{CB}  =
\sum_I \, \frac{\delta U(\Eone,\Etwo,  \ldots)}{\delta \EI_\mu^{~B}} \EI_\mu^{~C} \, \eta_{CA} \, .
\ee
Thus for a (connected) theory with $\N$ vielbeins, there are in fact only $\N-1$ independent Lorentz constraints.  The overall Lorentz invariance removes $D(D-1)/2$ components from the final vielbein.  This means $\N \times \tfrac{1}{2} D(D-1)$ components of all the $\N$ vielbeins can be eliminated, leaving the same number of components as the $\N$-metric theory.

These constraints play a crucial role in relating the vielbein theories to equivalent metric theories.  Consider the bi-vertex theory, given by \eqref{drgtvielbi}.  The potential is given by $U = \sum_n \beta_n \, U_n$, for generic coefficients $\beta_n$, where in matrix notation, the possible interaction terms are
\bea
U_0 &\!\!\!\!=\!\!\!\!& \det \Eone \, , \\ \nonumber
U_1&\!\!\!\!=\!\!\!\!& \det \Eone\, [\Eone^{-1}\Etwo]  \, , \\  \nonumber
U_2&\!\!\!\!=\!\!\!\!& \tfrac{1}{2}
\det \Eone\left( [\Eone^{-1}\Etwo]^2-[\Eone^{-1}\Etwo\,\Eone^{-1}\Etwo]\right) \, ,  \\ \nonumber
U_3&\!\!\!\!=\!\!\!\!& \tfrac{1}{6}
 \det \Eone \left([\Eone^{-1}\Etwo]^3-3[\Eone^{-1}\Etwo][\Eone^{-1}\Etwo\,\Eone^{-1}\Etwo]
 +2[\Eone^{-1}\Etwo\,\Eone^{-1}\Etwo\,\Eone^{-1}\Etwo] \right)  \, , \\ \nonumber
\vdots&&  
\eea

There are two vielbeins, so we expect one independent Lorentz constraint.  Using \eqref{c}, it is straightforward to determine this constraint to be
\be
\Eone^{-1}\Etwo\, \eta = \( \Eone^{-1}\Etwo \,\eta \)^T \, .
\ee
(By taking the inverse of both sides, we see that this is equivalent to $\Etwo^{-1}\Eone\, \eta = \( \Etwo^{-1}\Eone \,\eta \)^T \,$ so the constraint is in fact symmetric $1\leftrightarrow 2$.)  Note that it is precisely this constraint that was used in eq \eqref{symmvielbi} in passing from the vielbein formulation to the metric formulation of bi-gravity.

This constraint happens to be independent of the coefficients $\beta_n$ that appear in front of the various potential terms $U_n$.  This is a very nice property of the bi-vertex theory.  As long as one considers only tree graphs of spin-2's that interact through the bi-vertex interactions given above, the constraints are of this simple form.  Each line in the graph corresponds to one Lorentz constraint: in a tree graph with $\N$ vielbeins there are $\N-1$ lines and $\N-1$ constraints.  For a line that connects vielbein $\EI$ with vielbein $\EJ$ one has the corresponding constraint $\EI^{-1}\EJ\, \eta = (\EI^{-1}\EJ\, \eta)^T$.  This means that for every tree graph of bi-vertex interactions, there is an equivalent metric formulation of the theory, given simply by replacing $\EI^{-1}\EJ$ with $\sqrt{\gI^{-1}\gJ}$.

As soon as one closes a loop in the graph, however, things get more complicated.  The number of lines in the graph now exceeds the number of Lorentz constraints.  To see what happens, consider the theory whose graph is a triangle, and where each side only contains the interaction $U_1$ with a generic coefficient.   The potential is
\be
U=\alpha_1 \det \Eone\, [\Eone^{-1}\Etwo]
+\alpha_2 \det \Etwo\, [\Etwo^{-1}\Ethr]+\alpha_3 \det \Ethr\, [\Ethr^{-1}\Eone] \, ,
\ee
for generic coefficients $\alpha_2,\alpha_2,\alpha_3.$
The graph has three lines but only two Lorentz constraints, given by enforcing the symmetry of the following two matrices,
\bea
&\alpha_1 (\det \Eone)\, \Eone^{-1} \, \Etwo\, \eta - \alpha_2 (\det \Etwo)\, \Etwo^{-1} \, \Ethr\, \eta& ,  \\
&\alpha_2 (\det \Etwo)\, \Etwo^{-1} \, \Ethr\, \eta - \alpha_3 (\det \Ethr)\, \Ethr^{-1} \, \Eone\, \eta& .
\eea
The constraints now depend on the details and coefficients of the potential.  What's more, due to the form of the constraints, it's evident that the vielbein theory is no longer equivalent to the metric theory with the replacement $\EI^{-1}\EJ \rightarrow \sqrt{\gI^{-1}\gJ}$.  Thus, we suspect (but have not proven) that the theory of \cite{Khosravi:2011zi} has Boulware-Deser ghosts at the non-linear level.

One can perform a similar analysis of the tri-vertex,
\be
U = \det \Eone \left([\Eone^{-1}\Etwo][\Eone^{-1}\Ethr]-[\Eone^{-1}\Etwo\,\Eone^{-1}\Ethr] \right) \, .
\ee
There are again three vielbeins and thus two independent Lorentz constraints, given by enforcing the symmetry of the following two matrices,
\bea
&[\Eone^{-1}\Etwo]\,\Eone^{-1} \Ethr\,\eta  - \Eone^{-1}\Etwo\,\Eone^{-1} \Ethr\,\eta& ,\\
&[\Eone^{-1}\Ethr]\,\Eone^{-1} \Etwo\,\eta  - \Eone^{-1}\Ethr\,\Eone^{-1} \Etwo\,\eta& .
\eea
The theory graph contains more lines than constraints, so again, the constraints are complicated and dependent on the form of the potential.  We see as well for the tri-vertex  that, due to the form of the constraints, the vielbein theory is no longer equivalent to the metric theory with the replacement $\EI^{-1}\EJ \rightarrow \sqrt{\gI^{-1}\gJ}$.  It remains to be seen if there exists an equivalent metric formulation for the vielbein loop-graph theories and theory graphs containing higher point vertices, beyond the bi-vertex.

\bibliographystyle{utphys}
\addcontentsline{toc}{section}{References}
\bibliography{spin2arXiv3}

\end{document}